\documentclass[aps,pra,twocolumn, notitlepage, superscriptaddress, longbibliography]{revtex4-1}

\usepackage{xcolor}
\usepackage{graphicx}
\usepackage[utf8]{inputenc}

\usepackage{amsmath}
\usepackage{braket}
\usepackage{amssymb}
\usepackage{amsthm}
\usepackage{dsfont}
\usepackage{enumitem}
\usepackage{xr}

\newcommand{\cw}{\text{c}} 
\newcommand{\cc}{\text{cc}}
\newcommand{\cH}{\text{cH}}
\newcommand{\cV}{\text{cV}}
\newcommand{\ccH}{\text{ccH}}
\newcommand{\ccV}{\text{ccV}}
\newcommand{\ccD}{\text{ccD}}
\newcommand{\ccA}{\text{ccA}}

\newcommand{\diag}{\mathop{\mathrm{diag}}\nolimits}

\newcommand{\Cwp}[1]{\ensuremath{C_\mathrm{#1}}}
\newcommand{\Cqwp}{\Cwp{QWP}}
\newcommand{\Chwp}{\Cwp{HWP}}
\newcommand{\Ceom}{\Cwp{EOM}}
\newcommand{\Ccc}[1]{\ensuremath{C_{{#1}}}}
\newcommand{\CLL}{\Ccc{LL}}
\newcommand{\CAB}{\Ccc{AB}}

\newif\ifcolor
\colortrue
\def\CA#1{{\ifcolor\color{red}\fi#1}}
\def\CB#1{{\ifcolor\color{green!60!black}\fi#1}}
\def\CC#1{{\ifcolor\color{blue}\fi#1}}
\def\CD#1{{\ifcolor\color{brown}\fi#1}}

\newcommand{\upb}{Integrated Quantum Optics, Universit\"at Paderborn, Warburger Strasse 100, 33098 Paderborn, Germany}
\newcommand{\prague}{Faculty of Nuclear Sciences and Physical Engineering, Czech Technical University in Prague, B{\v r}ehov\'{a} 7, 115 19, Praha 1, Czech Republic}
\newcommand{\wigner}{Wigner Research Center for Physics, Hungarian Academy of Sciences, Konkoly-Thege M.\ u.\ 29-33, H-1121 Budapest, Hungary}

\begin{document}
\title{Photonic quantum walks with four-dimensional coins}

\author{Lennart Lorz} 
\affiliation{\upb}
\author{Evan Meyer-Scott} 
\affiliation{%
\upb
}%
\author{Thomas Nitsche}
\affiliation{\upb}
\author{V\'aclav Poto\v cek}
\affiliation{%
\prague
}%
\author{Aur\'el G\'abris}
\affiliation{\prague}
\affiliation{\wigner}
\author{Sonja Barkhofen} \email{sonja.barkhofen@uni-paderborn.de}
\affiliation{\upb}
\author{Igor Jex}
\affiliation{%
\prague
}%
\author{Christine Silberhorn}
\affiliation{\upb}

\begin{abstract}
The dimensionality of the internal coin space of discrete-time quantum walks has a strong impact on the complexity and richness of the dynamics of quantum walkers.
While two-dimensional coin operators are sufficient to define a certain range of dynamics on complex graphs, higher dimensional coins are necessary to unleash the full potential of discrete-time quantum walks.
In this work we present an experimental realization of a discrete-time quantum walk on a line graph that, instead of two-dimensional, exhibits a four-dimensional coin space.
Making use of the extra degree of freedom we observe multiple ballistic propagation speeds specific to higher dimensional coin operators.
By implementing a scalable technique, we demonstrate quantum walks on circles of various sizes, as well as on an example of a Husimi cactus graph.
The quantum walks are realized via time-multiplexing in a Michelson interferometer loop architecture, employing as the coin degrees of freedom the polarization and the traveling direction of the pulses in the loop.
Our theoretical analysis shows that the platform supports implementations of quantum walks with arbitrary $4 \times 4$ unitary coin operations, and usual quantum walks on a line with various periodic and twisted boundary conditions.
\end{abstract}

\maketitle

\section{Introduction}
During the past two decades, quantum walks \cite{aharonov_quantum_1993, meyer_absence_1996, farhi_quantum_1998}, the quantum mechanical analogue of random walks, have become an established basis for quantum algorithms \cite{childs_example_2002, shenvi_quantum_2003, ambainis_coins_2005, ambainis_quantum_2007, childs_universal_2013} and quantum simulations \cite{strauch_relativistic_2006, witthaut_quantum_2010, engel_evidence_2007, mohseni_environment-assisted_2008, lee_quantum_2015}.
Quantum walks (QWs) have been realized experimentally on various platforms, such as photons \cite{bouwmeester_optical_1999, perets_realization_2008, peruzzo_quantum_2010, broome_discrete_2010, schreiber_photons_2010, schreiber_2d_2012, sansoni_two-particle_2012, crespi_anderson_2013, cardano_quantum_2015, xue_localized_2015}, ions \cite{schmitz_quantum_2009, zahringer_realization_2010}, atoms \cite{genske_electric_2013, karski_quantum_2009, preiss_strongly_2015} and nuclear magnetic resonance \cite{du_experimental_2003}.
A detailed introduction to experimental implementations of quantum walks can be found in Ref.~\cite{wang_physical_2013}. 
Discrete time quantum walks (DTQWs) have been successfully implemented using time-multiplexing techniques \cite{schreiber_photons_2010, schreiber_2d_2012,  nitsche_quantum_2016, chen_observation_2018}, offering flexibility and easy reconfigurability accompanied by high efficiency and stability. 
A marked feature of the DTQW is an internal degree of freedom --- the coin space --- that conditions the spatial shift of the walker, in the same way as a coin toss determines the movement of a classical random walker.
It is the dynamics in the coin space that is argued to provide the key ingredient to the complex behavior of the DTQW \cite{ambainis_coins_2005}.

While the initial definition of DTQWs assumed translation invariant and time independent dynamics, more versatility can be obtained by spatial and temporal control of the quantum walk parameters.
By varying the coin operation such systems have been used experimentally to observe Anderson localization \cite{schreiber_decoherence_2011, crespi_anderson_2013, xue_localized_2015}, dynamical localization \cite{genske_electric_2013}, topological phases \cite{kitagawa_observation_2012,  rechtsman_photonic_2013, poli_selective_2015, zeuner_observation_2015, cardano_statistical_2016, xiao_observation_2017, cardano_detection_2017, barkhofen_measuring_2017, wang_experimental_2018}, and other fundamental effects such as recurrence \cite{nitsche_probing_2018} and revivals \cite{xue_experimental_2015}.
The dynamic control of the coin operation can be extended to engineering the topology of the graph on which the walk takes place: finite \cite{nitsche_quantum_2016} and percolation graphs \cite{elster_quantum_2015}, and lines with periodic boundary conditions \cite{bian_experimental_2017} have been demonstrated experimentally.

To have any effect on the walker dynamics, the minimum required dimensionality for the coin space is two.
In order to reduce the required theoretical and experimental effort associated with the study of higher dimensional coins, many of the above works employ multi-step protocols, which use only two-dimensional coins.
These protocols simulate higher dimensional coins by splitting up each step into multiple coin and shift operations acting on a two-dimensional coin space.
They have found use not only in realizing dynamics on graphs embedded in higher dimensions, but also in 1D quantum walks on more sophisticated graphs, such as on percolation graphs or circles \cite{elster_quantum_2015, bian_experimental_2017}.
However, as the required doubling or even triplication of the necessary step numbers for the implementation of such multi-step schemes is experimentally disadvantageous in terms of losses, inaccuracies, and scalability, these protocols significantly impact the efficiency of the physical implementation.

Already on the one dimensional (1D) line DTQWs with higher dimensional coins have been shown to exhibit unique features not possessed by two-dimensional coins, among the most striking the so-called trapping \cite{inui_one-dimensional_2005, stefanak_stability_2014, dan_one-dimensional_2015}.
While due to the simplicity of the 1D structure these may be regarded as toy systems, they can be efficiently used to demonstrate several fundamental differences between classical and quantum walks.
Trapping can be used for instance in conjunction with dynamically controlled coin operators to shape the profile of the walker's wave packet, having no counterpart in classical random walks.
In the case of more complex graph topologies (e.g.\ graphs embedded in higher dimensional spaces, or nonregular graphs) the dimensionality of the coin space will provide the critical ingredient for more involved or even unexpected applications.
For example, DTQWs with genuine four-dimensional coins on structures embedded in the two-dimensional (2D) space admit phases analogous to the quantum spin Hall (QSH) phases \cite{kitagawa_exploring_2010, kane_quantum_2005}, offering significantly new applications over the phases accessible in 1D \cite{kitagawa_exploring_2010, tarasinski_scattering_2014}.
Another example is that of the Grover walk on a 2D grid, exhibiting dynamics composed of a spreading and a localized part \cite{tregenna_controlling_2003, inui_localization_2004}, of which only the spreading part can be reproduced by two-dimensional coins \cite{di_franco_mimicking_2011}.
These limitations of two-dimensional coins provide a strong motivation to achieve efficient implementations of quantum walks with genuine higher dimensional coin operators while maintaining precise dynamic control.
While there have been theoretical proposals \cite{hamilton_quantum_2011, neves_photonic_2018}, and limited experimental realizations of higher dimensional coins \cite{zahringer_realization_2010, schreiber_2d_2012, xue_localized_2015, derrico_two-dimensional_2018}, no universal scalable platform has been demonstrated yet.
\begin{figure}
	\includegraphics[width=.9\columnwidth]{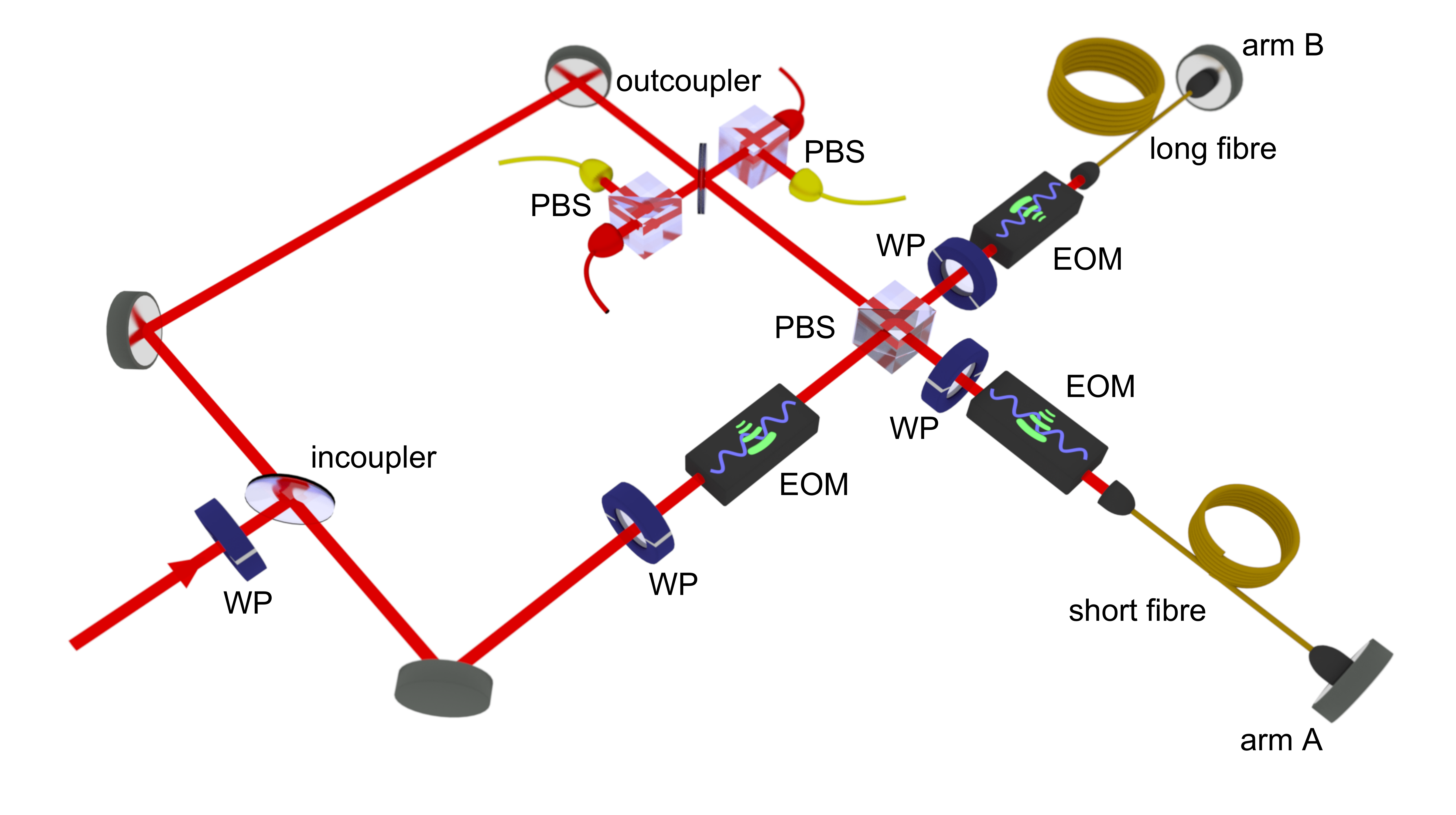}
	\caption{Experimental realization of the Michelson geometry. The three $2\times 2$ polarization rotations forming the coin are realized by electro-optic modulators (EOM) in combination with either half or quarter waveplates (WP). We use single mode fibers of 328\,m and 338\,m length in the arms; the other parts are in free space. A waveplate in front of the incoupling mirror determines the input polarization of the pulse. After the outcoupling at the partially reflective mirror the pulses are routed to four superconducting nanowire single photon detectors enabling the resolution of all four internal states.}
	\label{fig:Setup}
\end{figure}
\begin{figure*}
	\includegraphics[width=0.8\textwidth]{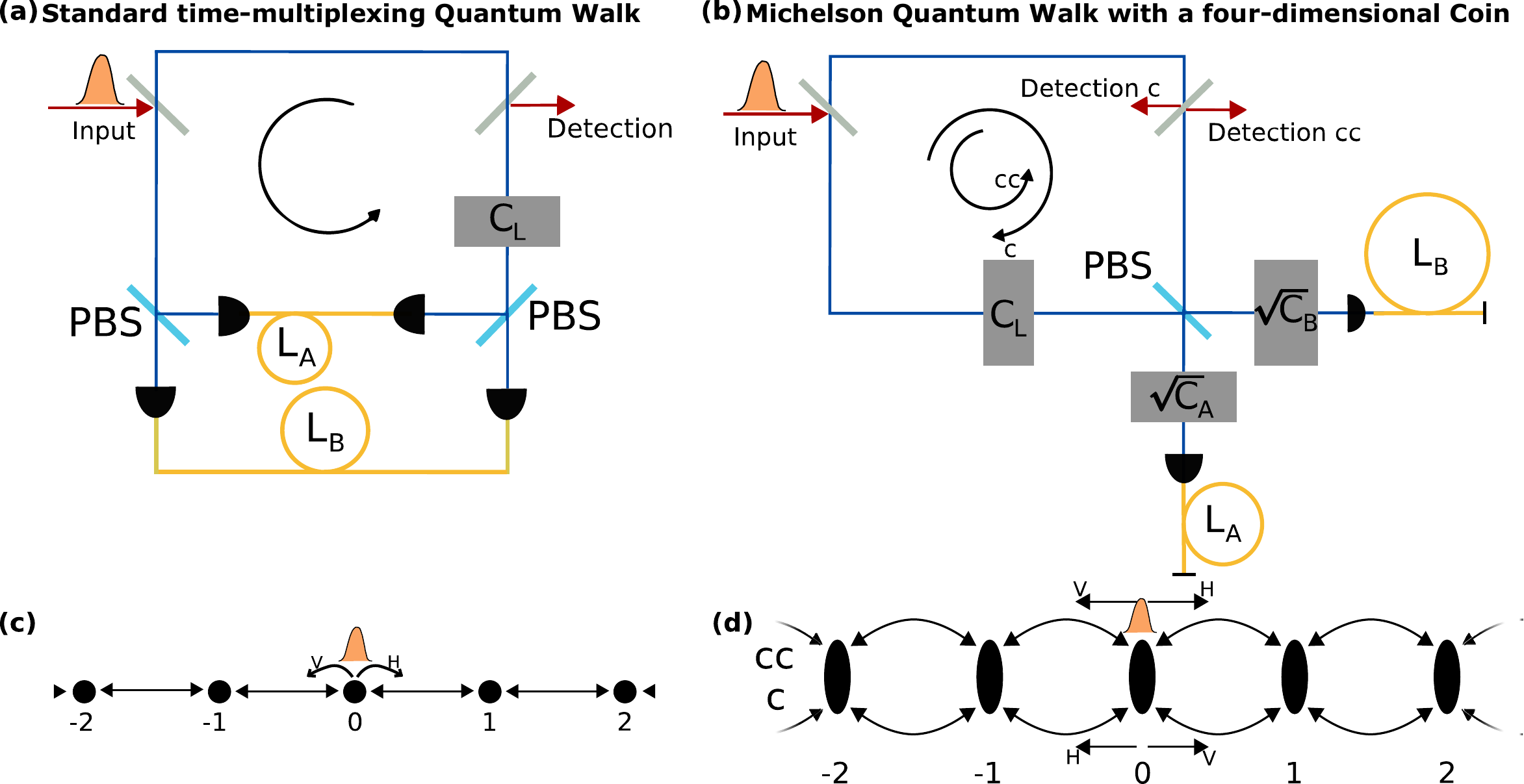}
	\caption{Schemes for the Mach--Zehnder-type (a) and for the Michelson-type loop geometry (b). In both cases the walker is coupled into the loop via a partially reflective beam splitter and is polarization rotated in the loop by optical elements realizing the operator $C_L$. At the polarizing beam splitter (PBS) the pulse is split according to its polarization and the two arms A and B with different lengths $L_A$ and $L_B$ introduce a well-defined time-delay between the constituents. In the Mach--Zehnder interferometer the pulses travel in the counter-clockwise direction, while in the Michelson geometry, both clockwise (denoted $\cw$) and counter-clockwise ($\cc$) traveling directions are used. The optical elements placed in the arms of the Michelson-type geometry realize the rotations $C_A$ and $C_B$ after the double passage. As a consequence, the walker can be characterized by four internal states (the two polarizations and two traveling directions) in addition to its current location. Graph representations of the 1D quantum walks by the Mach--Zehnder and Michelson-type geometries, respectively, are shown in (c) and (d), illustrating the role of the different polarization and propagation modes.}
	\label{fig:schemes}
\end{figure*}

In this work we present experimental implementations of DTQWs on a line governed by programmably controlled four-dimensional coins, reaching beyond the previous two-dimensional definition and demonstrating QWs on new complex graph topologies.
At the heart of our time-multiplexing scheme is an interferometer arranged in a Michelson-type geometry, in contrast to earlier implementations based on a Mach--Zehnder geometry.
While offering identical stability and versatility, the present setup introduces a new degree of freedom for the coin, namely the direction of propagation of two counter-propagating optical modes.
Combining these with polarization supports the four-dimensional coin.
The higher-dimensional coin space and the temporal control of the coin operations 
enable us to efficiently realize DTQWs on non-trivial graphs of different sizes and topologies.

The structure of the paper is as follows. In Sec.~\ref{sec:exp-apparat} we introduce the experimental apparatus we use for realizing quantum walks with four-dimensional coins, point out its differences to the earlier time-multiplexing setup and detail the principle of operation.
Sec.~\ref{sec:theo} formalizes the DTQW time evolution of the Michelson geometry and analyzes the attainable dynamics.
In Sec.~\ref{sec:exp-results} we present experimental results on the realization of quantum walks on various graph structures.
First, we demonstrate the usual Hadamard quantum walks by restricting the dynamics into an invariant two-dimensional subspace with a suitable choice of the four-dimensional coin operator.
Next, we realize a quantum walk with a genuine four-dimensional coin and observe the emergence of multiple lobes in the probability distribution, characteristic of translationally invariant DTQWs with higher dimensional coins.
By dynamically controlling the coin, we extend 1D DTQWs with Hadamard and non-mixing coins onto circles of various sizes, and demonstrate effects of periodic boundary conditions, in particular the equidistribution of the quantum walker.
Finally, we present a quantum walk on a minimal example of a Husimi cactus graph consisting of a pair of connected circles, resembling a figure eight, with the connecting node characterized by a four-dimensional operator.
We discuss the significance of the results and provide an outlook in Sec.~\ref{sec:discussion}.

\section{Experimental apparatus}
\label{sec:exp-apparat}
The layout of our experiment, depicted in Fig.~\ref{fig:Setup}, resembles a Michelson interferometer closed by a loop.
The coherent laser pulse (wavelength 1550 nm) plays the role of the quantum walker, using the mathematical equivalence between wave dynamics and single particle quantum dynamics \cite{paul_introduction_2004}.
The input pulse is coupled into the loop by a beam splitter with low reflectivity $R \approx 1\,\%$ ensuring high transmittivity for the traveling pulses.
The loop allows two propagation directions, clockwise (c) and counter-clockwise (cc), and for each direction we can distinguish two orthogonal polarizations, horizontal (H) and vertical (V) (cf. Fig.~\ref{fig:schemes}b).
We label these four orthogonal modes by $\cH$, $\cV$, $\ccH$ and $\ccV$. 
To control the dynamics of the pulses, we insert polarization rotating elements consisting of waveplates and fast-switching electro-optical modulators (EOMs) in the arms $A$ and $B$ as well as in the loop (cf. Fig.~\ref{fig:Setup}).
The initial pulse with a well defined polarization is coupled into the modes $\ccH$ and $\ccV$ by the incoupler.
The polarization of the pulse is rotated by the waveplate and the EOM before it reaches the polarizing beam splitter (PBS) and enters the arms $A$ and $B$.
After a reflection in the arms the pulse re-enters the loop and is split into the four available modes, depending on the arm's polarization rotation. 
For detection of the pulses, we place another weakly reflecting beam splitter ($R \approx 2\,\%$) in the loop and use a pair of PBSs and four superconducting nanowire single-photon detectors to discriminate the four internal states.
By using single mode optical fibers of different lengths (328 and 338\,m) in the arms $A$ and $B$, we can introduce a well-defined time delay of $\tau_\mathrm{pos} = 95$\,ns between pulses that took different arms.
By choosing the time delay to be longer than the pulse widths ($\approx 100$\,ps) and detector dead times ($\approx 90$\,ns with 90\,\% recovery of the efficiency), we can resolve the outcoupled pulses with different delays and associate them to unique time bins.
The roundtrip efficiency (i.e. the transmission from one step to another) in the looped interferometer is $63\pm3$\,\%, which significantly improves the performance of earlier setups with efficiencies of $\approx 40$\,\% as presented in \cite{nitsche_quantum_2016}.
In order to achieve a good signal-to-noise ratio for high step numbers we perform measurements with two different initial power levels, which are then concatenated.
This concatenation of two data sets is necessary since for a low power input the signal becomes too small after a small number of steps, while the high input powers cause detector saturation for the early steps and make a reliable probability extraction impossible.
In each case we normalize the total intensity per step to one which is then equivalent to the walker's probability distribution.

In Fig.~\ref{fig:schemes} we illustrate the dynamics of the interferometer, based on the time-multiplexing technique. For reference we additionally provide the Mach--Zehnder-type geometry for the visualisation of the standard principle of time-multiplexing quantum walks as detailed in \cite{schreiber_photons_2010, nitsche_quantum_2016}.
To understand the dynamics of the Michelson interferometer, it is instructive to follow what happens to pulses coming from the loop, impinging on the PBS in all four modes $\cH$, $\cV$, $\ccH$ and $\ccV$ at once.
The PBS guides the pulses from modes $\cH$ and $\ccV$ into arm $A$, and from modes $\ccH$ and $\cV$ to arm $B$.
In each arm the polarized pulses are rotated by optical elements implementing $C_A$ and $C_B$ respectively, and a relative time delay between the two different paths is introduced.
Back at the PBS, the pulses are reflected or transmitted according to their polarization, such that e.g.\ the originally horizontal and clockwise traveling pulse, upon entering and leaving arm $A$, is mapped onto modes $\ccH$ and $\cV$ for the next loop iteration.
A full roundtrip is thus defined by a rotation of the clockwise and counter-clockwise propagating pulses by the elements in the loop, followed by the mode-dependent rotation and delay in the two arms $A$ and $B$.

Particularly simple dynamics can be observed if the optical elements in the arms $A$ and $B$ are set up such that the net effect of the double passage and reflection is a rotation of the pulse polarization by 90$^{\circ}$.
In this case an initially counter-clockwise traveling pulse continues to travel in the counter-clockwise direction after returning from the arms $A$ and $B$.
Here the only role of the arms $A$ and $B$ is to provide a polarization dependent delay, while mixing of polarizations depends solely on the elements located inside the loop.
By controlling the elements inside the loop, a wide range of general 1D quantum walk dynamics is accessible --- limited only by the capabilities of the available optical components.

\section{Mathematical description}
\label{sec:theo}
\subsection{Time evolution of pulses as a quantum walk}
For the purposes of mathematical description, we use a formal mapping between a wave mechanical superposition of spatially or temporally separated optical pulses and a quantum mechanical superposition of states of a photon \cite{paul_introduction_2004} representing the quantum walker, as employed in our previous works \cite{schreiber_photons_2010, schreiber_2d_2012, nitsche_quantum_2016, nitsche_probing_2018}.

The state of a discrete-time quantum walker is described by $\ket{\Psi}$, a vector in the corresponding tensor product Hilbert space $\mathcal{H} = \mathcal{H}_c \otimes \mathcal{H}_x$.
For a DTQW on a line, the position Hilbert space $\mathcal{H}_x$ equals $l^2(\mathbb{Z})$, spanning all possible positions $x$ associated with the basis vectors $\{ \ket{x} \mid  x \in \mathbb{Z} \}$. 
The coin Hilbert space, $\mathcal{H}_c$, describes the internal degree of freedom.
For a 1D walk a two-dimensional coin space is usually assumed, which facilitated the use of polarization for this purpose by a number of research groups (see e.g.\ \cite{broome_discrete_2010, schreiber_photons_2010, sansoni_two-particle_2012, xue_experimental_2015}).
In a Michelson geometry (Fig.~\ref{fig:schemes} (b)) the walker can additionally be in a superposition of the two traveling directions in the loop, resulting in a four-dimensional coin space for a 1D walk.
We introduce four orthogonal basis states $\mathcal{H}_c$ as $\{ \ket{\cH},  \ket{\cV},  \ket{\ccH},  \ket{\ccV} \}$ representing the four orthogonal modes introduced earlier.  
\begin{eqnarray} \label{eq:state}
	\ket{\Psi} &=& \sum\limits_{x\in\mathbb{Z}} \big(\alpha_{\cH,x}\ket{\cH} \otimes \ket{x} 
	+ \alpha_{\cV,x}\ket{\cV} \otimes \ket{x} \\ \nonumber
	&&{}+ \alpha_{\ccH,x}\ket{\ccH} \otimes \ket{x} + \alpha_{\ccV,x}\ket{\ccV} \otimes \ket{x}\big)
\end{eqnarray}
with the complex coefficients $\alpha_{d,x}$ obeying $\sum_d \sum_x |\alpha_{d,x}|^2 = 1$.

The unitary evolution of a DTQW is determined by the coin operator $\hat{C}$ acting on the internal degree of freedom, followed by the step operator $\hat{S}$, which performs a conditional shift in the position $x$; together we write $ \ket{\Psi_{t+1}} = \hat{S} \hat{C} \ket{\Psi_t}$.
In the convention defined by the experimental setup (Fig.~\ref{fig:schemes} (b)),
$\hat{S}$ shifts the basis states $\ket{\cH}$ and $\ket{\ccV}$ ($\ket{\ccH}$ and $\ket{\cV}$) one position to the left (right), which corresponds to earlier (later) arrival times, and simultaneously reverses the traveling direction;
in quantum walk terminology such conditional shift combined with a reverse in direction is commonly referred to as a flip-flop step operator.
Formally, the operator can be expressed as
\begin{equation}
  \begin{aligned}
    \hat{S} = &\sum_{x} 
    \big(\ket{\ccH}\!\bra{\cH} \otimes \ket{x-1}\!\bra{x}
    + \ket{\ccV}\!\bra{\cV} \otimes \ket{x+1}\!\bra{x} \\
    &+ \ket{\cH}\!\bra{\ccH} \otimes \ket{x+1}\!\bra{x} 
    + \ket{\cV}\!\bra{\ccV} \otimes \ket{x-1}\!\bra{x}\big).
  \end{aligned}
\label{eq:stepoperator}
\end{equation}
Since the position space is still one dimensional but two different coin states indicate a step to the left and two to the right, the structure of the walk can be visualized as a line graph with doubled edges as illustrated in Fig.~\ref{fig:schemes} (d).

The coin matrix describes the combined action of three $2\times 2$ polarization rotations defined by the three operations $C_L$, $C_A$ and $C_B$ in the loop and the two arms, respectively.
Note that the elements in the arms $A$ and $B$ are passed twice by each pulse entering the respective arm; by $C_A$ and $C_B$ we describe the full rotation accumulated by the time it re-enters the loop.
To realize the desired polarization rotations, we use quarter-wave plates (QWPs), half-wave plates (HWPs) and EOMs.
In the polarization basis $\{ \ket{\mathrm{H}}, \ket{\mathrm{V}}\}$ the waveplates aligned at an angle $\alpha$ are characterized by the matrices
\begin{equation} \label{eq:QWP}
\Cqwp(\alpha) = \frac{-i}{\sqrt{2}} \begin{pmatrix} \cos2\alpha+i & \sin2\alpha \\ \sin2\alpha &-\cos2\alpha+i
\end{pmatrix},
\end{equation}
and
\begin{equation} \label{eq:HWP}
  \Chwp(\alpha) = \begin{pmatrix} \cos2\alpha & \sin2\alpha \\
    \sin2\alpha & -\cos2\alpha
\end{pmatrix},
\end{equation}
respectively.
The EOMs are aligned such that they are described by matrices
\begin{equation} \label{eq:EOM}
  \Ceom(\varphi) = \begin{pmatrix} \cos\varphi & -i\sin\varphi \\
    -i\sin\varphi & \cos\varphi
\end{pmatrix},
\end{equation}
with the phase $\varphi$ depending on the voltage applied to the particular EOM during a particular time bin \cite{nitsche_quantum_2016}. In all cases involving dynamic EOM switches we always align the quarter- and half-wave plates at $\alpha = 45^\circ$, so that the matrices \eqref{eq:QWP} and \eqref{eq:HWP} commute with \eqref{eq:EOM} and it is inconsequential in which order a pulse encounters them. 

Since the elements in the loop do not mix counter propagating pulses, their effect can be described in the basis $\{ \ket{\cH},  \ket{\cV},  \ket{\ccH},  \ket{\ccV} \}$ by the block diagonal matrix
\begin{equation}
\CLL = \begin{pmatrix}
C_{L,HH} & C_{L,HV} & 0 & 0 \\
C_{L,VH} & C_{L,VV} & 0 & 0 \\
0 & 0 & C_{L,HH} & C_{L,HV} \\
0 & 0 & C_{L,VH} & C_{L,VV} 
\end{pmatrix}.
\label{eq:C_LL}
\end{equation}
Due to the action of the PBS, the optical elements in the arms $A$ and $B$ mix pulses from different traveling directions, so that the total operation corresponds to the matrix
\begin{equation}
\CAB = \begin{pmatrix}
C_{A,HH} & 0 & 0 & C_{A,HV} \\
0 & C_{B,VV} & C_{B,VH} & 0 \\
0 & C_{B,HV} & C_{B,HH} & 0 \\
C_{A,VH} & 0 & 0 & C_{A,VV} \\
\end{pmatrix},
\label{eq:C_AB}
\end{equation}
transforming e.g.\ $\ket{\cH}$ into a superposition of $\ket{\cH}$ and $\ket{\ccV}$.
The coin matrix of the quantum walk arises as the product of these two matrices
\begin{equation}
\label{eq:fullCoin}
C = \CAB \CLL.
\end{equation}

When the polarization rotations are static in time, we can express the full coin operator as $\hat{C} = C \otimes \mathds{1}_x$.
However, due to the unique relation between time bins and position and step number of the walker, we can program specific phase shifts $\varphi_{t,x,A}$, $\varphi_{t,x,B}$ and $\varphi_{t,x,L}$ to be realized for each time bin $x$ by the three EOMs, thus making the coin operator position and time dependent, formulated as $\hat{C}_t = \sum_x C_{t,x} \otimes \ket{x}\!\bra{x}$. In this work we will only make use of the position dependence of the coin, keeping the same operations for each step $t$.

\subsection{Set of directly accessible coins, achieving universality with a three-step protocol}

The product \eqref{eq:fullCoin} covers a useful subset of $U(4)$, as demonstrated by the experiments described in the following sections.
Suppose, we intend to realize a certain coin operator, how to tell if this target coin can be decomposed into this form?
It turns out that there is a particularly simple condition, requiring the pairwise linear independence of two appropriately chosen pair of vectors formed from the elements of the $4\times4$ coin matrix.
We have included the proof in \ref{sec:coins-one-roundtrip}.
The proof is constructive in the sense that it shows how to efficiently find a decomposition Eq.~(\ref{eq:fullCoin}) for a particular coin $C$, provided it exists.

A larger class of coins can be covered by using operators $\CLL'$ without the restriction that they act the same on $\cw$ and $\cc$-propagating pulses.
This can be achieved e.g.\ by altering our setup such that counter-propagating pulses reach the loop EOM with a sufficient time difference, allowing the programming of different rotations. 

The full $U(4)$ can be recovered by employing a multi-step protocol \cite{kitagawa_observation_2012, elster_quantum_2015, bian_experimental_2017} consisting of three steps.
A crucial fact that the protocol uses is that any four-dimensional unitary matrix can be written as a product of two matrices each of the form $\CAB \CLL'$ -- this we prove rigorously in \ref{sec:coins-universality}.
Leveraging on the flip-flop nature of $\hat{S}$, namely that two successive applications of the step operator cancel each other $\hat{S}\cdot\hat{S} = \hat{\mathds{1}}$, we consider a sequence of coins $C_1$, $\mathds{1}$ and $C_2$ leading to an overall transformation described by $\hat{S} \hat{C_2} \hat{C_1}$.
Appendix \ref{sec:coins-universality} contains additional details of these arguments.
We note, that besides static coins of the form $\hat{C} = C \otimes \mathds{1}_x$, the protocol is applicable also to position and time dependent distributions.

\subsection{Dynamical features of four-dimensional coins}
\label{sec:theo_gv}
Calculating analytically the evolution of a quantum walker over many steps is generally a demanding task.
However, for translationally invariant systems it is possible to characterize the long time asymptotic dynamics in a simple way by analyzing the dispersion relation, i.e.\ the $k$-dependent quasi-energies $\omega(k)$ of the unitary evolution operator obtained after performing the spatial Fourier transform \cite{ambainis_one-dimensional_2001, hinarejos_understanding_2013}.
By locating all local extrema of the group velocities defined as the derivative $v_g(k) = d\omega(k)/dk$ we can determine the number and propagation speeds of wavefronts emerging from an initially localized state.
In the case of a standard 1D quantum walk with a two-dimensional coin, this analysis yields the well-known double-lobed position distribution (see e.g. Fig.~\ref{fig:HadQW}), with the two wavefronts moving away from the origin at speeds equal to the absolute value of the diagonal elements of the coin matrix, i.e. $\pm 1/\sqrt{2}$ for the Hadamard walk.
While split-step walks exhibit a richer dynamics in many respects, their asymptotic dynamics is still characterised by a double-lobed distribution owing to the similarity of their dispersion relation with the standard 1D walk (see \ref{sec:ss-disp}  for details).
DTQWs with higher dimensional coins, however, have been shown to feature additional ballistically propagating or trapped wavefronts \cite{stefanak_continuous_2012}.
We would like to note that the simple analysis of the dispersion relation cannot account for the effect of the initial coin state, which generally influences the relative intensities of the ballistic wavefronts, and neither does it provide a characteristic time after which the asymptotic dynamics is guaranteed to set in.

\begin{figure}
	\includegraphics[width=.9\columnwidth]{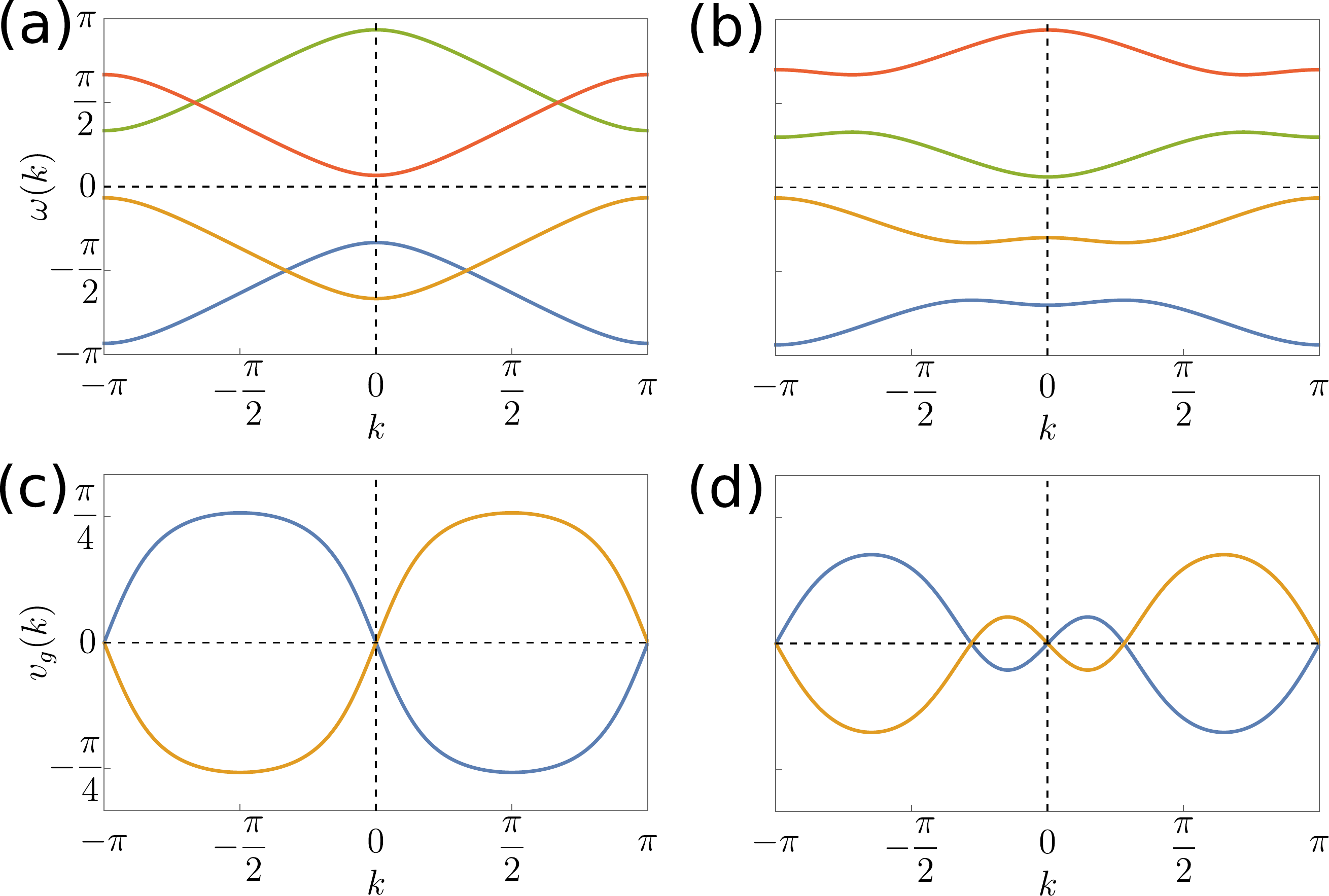}
	\caption{Quasi-energies $\omega(k)$ as functions of $k$ (up to irrelevant phase factors) for quantum walks on infinite one dimensional lines with the coin operators being a) $C_A= C_B= \Cqwp(12^{\circ})$, $C_L = \Chwp(27^{\circ})$, and b) $C_A= C_B= \Cqwp(27^{\circ}) \Chwp(0^{\circ}) \Cqwp(27^{\circ})$, $C_L= \Chwp(20^{\circ})$. The first example exhibits level crossings, the second example features level repulsions, and together illustrate how avoided crossings result in additional bends and inflection points. Subfigures c) and d) show the group velocities $v_g(k)=d\omega(k)/dk$ calculated from two branches of the spectra plotted on a) and b), respectively. The local extremal points of $v_g(k)$ yield the experimentally observable wavefront velocities.}
	\label{fig:dispersion-4D}
\end{figure}

We have found that four-dimensional coins can give rise up to eight wavefronts in the position distribution of the walker, see \ref{sec:4d-disp} for additional remarks.
When the coin operators $C_A$, $C_B$ and $C_L$ are restricted to quarter- and half-wave plates the symmetries of the system permit degeneracies allowing crossings between different quasi-energy branches, as illustrated on Fig.~\ref{fig:dispersion-4D}a.
Under these restrictions, we observe a behavior similar to a standard 1D walk, exhibiting the usual double-lobe distribution.
By considering coin operators built up from several waveplates, we can lift these degeneracies and turn the level crossings between the branches of quasi-energies into avoided crossings (see Fig.~\ref{fig:dispersion-4D}b).
The level repulsion introduces additional bends and thus additional inflection points to the dispersion curves.
The new inflection points can give rise to additional wavefronts associated with each distinct propagation velocities.
The particular propagation velocities can be controlled by the appropriate choice of the coin operator.
With the correct choice, the different propagation speeds may be discerned even within a limited number of steps of the QW evolution.

\section{Experimental results}
\label{sec:exp-results}
The experimental results we present in this section can be divided into two groups.
The experiments reported in Secs.~\ref{sec:HadWalks} and \ref{sec:Full4dWalks}
explore the translationally invariant dynamics of a quantum walker with a four-dimensional coin operator, using a (static) WP to implement the coin operators $\CLL$ and $\CAB$.
In Secs.~\ref{sec:CircleWalks} and \ref{sec:EightWalks} we study dynamics on finite cyclic graphs of various topologies, using three dynamically controlled EOMs as shown on Fig.~\ref{fig:Setup}.

\subsection{Hadamard walk}
\label{sec:HadWalks}
\begin{figure}
	\includegraphics[width=.95\columnwidth]{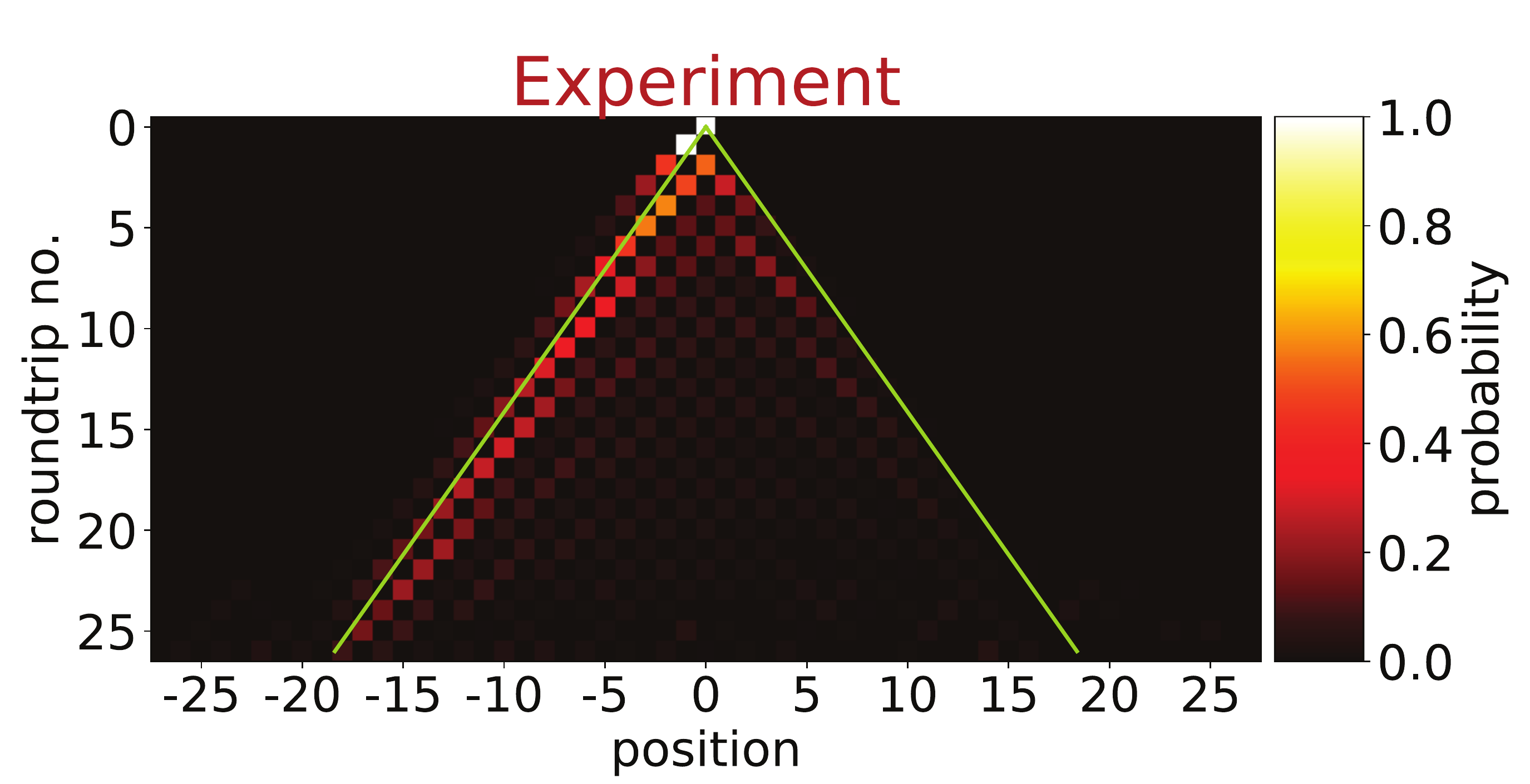}
	\caption{Experimentally obtained intensity evolution in the $\cc$ travelling direction of a Hadamard walk initialized in $\ket{\ccA}$, after summation over the polarization degree of freedom.
The intensities in the $\cw$-direction are zero. 
A pair of solid lines indicate the wavefront trajectories at the speeds of $\pm 1/\sqrt{2}$ expected from the asymptotic analysis of Sec.~\ref{sec:theo_gv}.
The expected double peak structure and propagation speeds are clearly observable within the experimentally attainable steps.}
	\label{fig:HadQW}
\end{figure}
To demonstrate the coherence properties of the setup in a simple manner, we present dynamics equivalent to a conventional DTQW on a line with a two-dimensional coin.
By appropriate choices of the initial state and the waveplate parameters we restrict the dynamics to an invariant subspace corresponding to the coin states $\ket{\ccH}$ and $\ket{\ccV}$, representing a conventional DTQW in the $\cc$ travelling direction of light pulses.
In particular, we set $C_L$ to a Hadamard operation, realized up to a global phase by a HWP at $\alpha = 22.5^\circ$, yielding
\begin{equation}
\CLL = \frac{1}{\sqrt{2}}\begin{pmatrix}
1 & 1 & 0 & 0 \\
1 & -1 & 0 & 0 \\
0 & 0 & 1 & 1 \\
0 & 0 & 1 & -1 
\end{pmatrix}.
\label{eq:C_LL_Had}
\end{equation} 
The polarization rotations $C_A$ and $C_B$ in the arms are set to a polarization swap by introducing QWPs at the angle $\alpha=45^\circ$ which are passed twice.
Thus, up to an irrelevant global $-i$ phase, the corresponding coin operator is
\begin{equation}
\CAB = \begin{pmatrix}
0 & 0 & 0 &1 \\
0 & 0 & 1 & 0 \\
0 & 1 & 0 & 0 \\
1 & 0 & 0 & 0 \\
\end{pmatrix}.
\label{eq:C_AB_Swap}
\end{equation}
This maps $\ket{\ccH}$ to $\ket{\cV}$ and $\ket{\ccV}$ to $\ket{\cH}$, and the subsequent step operator \eqref{eq:stepoperator} brings the traveling direction back to $\cc$.
Due to the absence of mixing of traveling directions, the pulses only ever travel in the loop in the counter-clockwise direction, in which the walk was initiated.

The system reduces to a quantum walk with a two-dimensional coin, and can be described by the effective step and coin operators as
\begin{eqnarray} \nonumber
\hat{S}_2 &=& \sum_{x\in\mathds{Z}} \big(\ket{R}\!\bra{R}\otimes \ket{x+1}\!\bra{x} + \ket{L}\!\bra{L} \otimes \ket{x-1}\!\bra{x} \big) \\ \label{eq:effCS}
C_2 &=&  \frac{1}{\sqrt{2}}\begin{pmatrix}
1 & 1  \\
1 & -1 
\end{pmatrix},
\end{eqnarray}
where we use $\ket{R}$ and $\ket{L}$ to follow the conventional notation for the right and left shifted components, respectively.
The abstract states $\ket{R}$ and $\ket{L}$ correspond to $\ket{\ccH}$ and $\ket{\ccV}$ in the experiment.

As a figure of merit we use the polarization resolved \textit{similarity} between experimental and numerical probabilities defined as
\begin{equation}\label{eq:sim}
\mathcal{S}(t) = \Big| \sum_{d,x} \sqrt{P_{d,x}^{(\mathrm{exp})}(t) P_{d,x}^{(\mathrm{num})}(t)} \Big|^2 ,
\end{equation}
for the relevant positions $x$ and the coin states $d$ at a certain step $t$.
We also make use of \textit{average similarity} (over $T$ steps) defined as $\bar{\mathcal{S}}=\frac1T \sum_{t=1}^{T} \mathcal{S}(t)$. 

The measured standard Hadamard walk over $T=25$ steps exhibits similarity of $\bar{\mathcal{S}}= 91.2\,\%$ to the theoretical expectation (Fig.~\ref{fig:HadQW}), demonstrating the outstanding coherence properties within the polarization degree of freedom of each propagation direction.

To test the robustness of coherence between the two propagation directions  additional measurements were performed, where we have implemented dynamics alternating between the $\ket{\cc}$ and $\ket{\cw}$ associated subspaces at every DTQW iteration (see \ref{sec:parrev} for the details).
The obtained similarity of $\bar{\mathcal{S}}=93.1\,\%$ between numerical and experimental data confirmed the high overall coherence properties of the setup indicating a good basis for implementing more advanced quantum walk dynamics.

\subsection{Walk with a genuine four-dimensional coin}
\label{sec:Full4dWalks}
\begin{figure}
\includegraphics[width=.99\columnwidth]{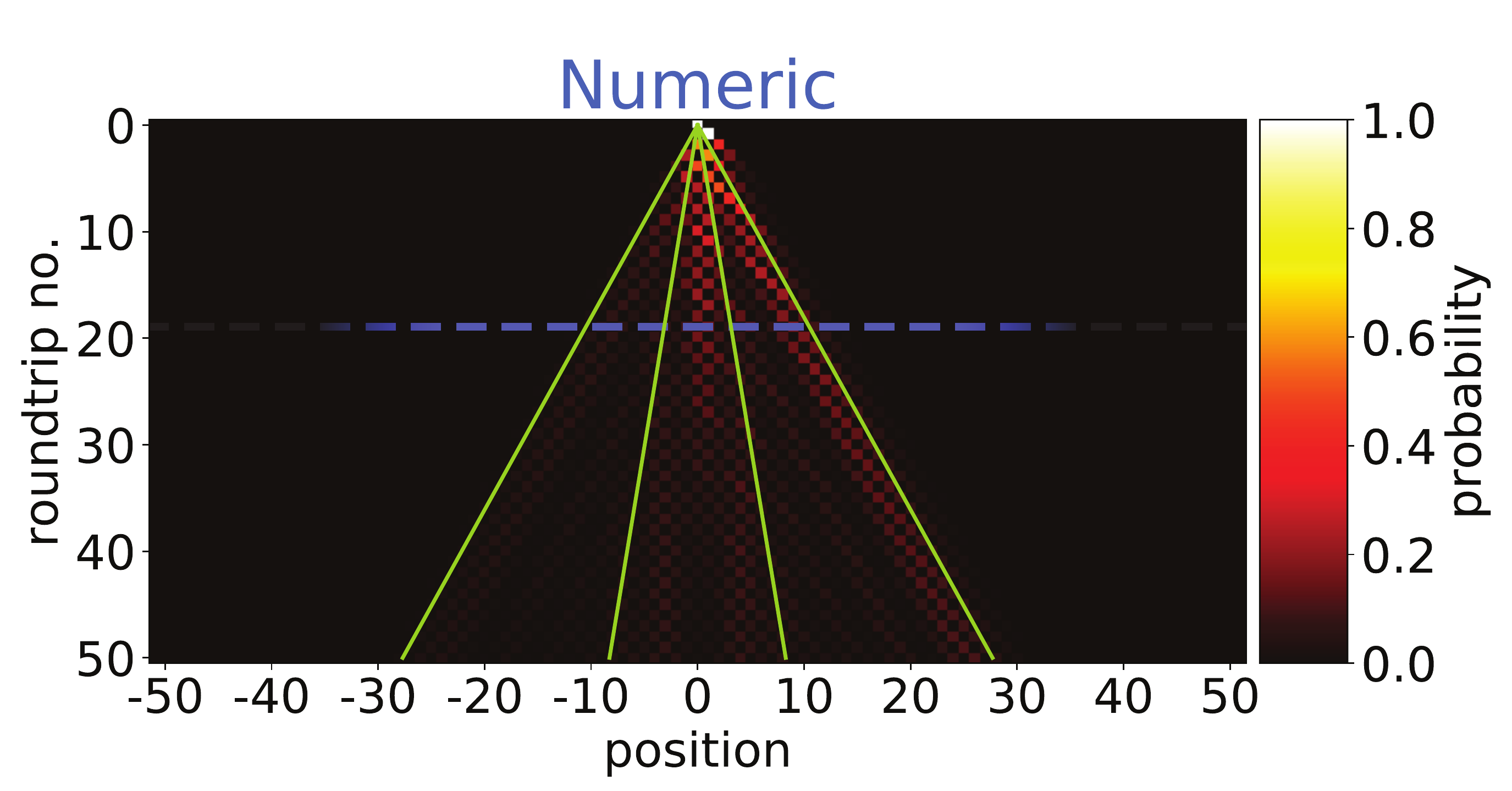}
\caption{Numerically calculated intensity distributions for the walk corresponding to Fig.~\ref{fig:dispersion-4D}b started from the initial state $\ket{\ccD}$, with all internal degrees of freedom summed up.
The solid lines have slopes $\pm0.1655$ and $\pm0.5538$ corresponding to the predicted asymptotic velocities of the wavefronts, and the dashed line indicates our experimental limit on the number of steps.
The plot confirms the validity of the asymptotic results for longer times, and also indicate a transient regime longer than for the Hadamard walk (see Fig.~\ref{fig:HadQW}).
}
\label{fig:groupVel-numeric}
\end{figure}

\begin{figure}
  \includegraphics[width=.9\columnwidth]{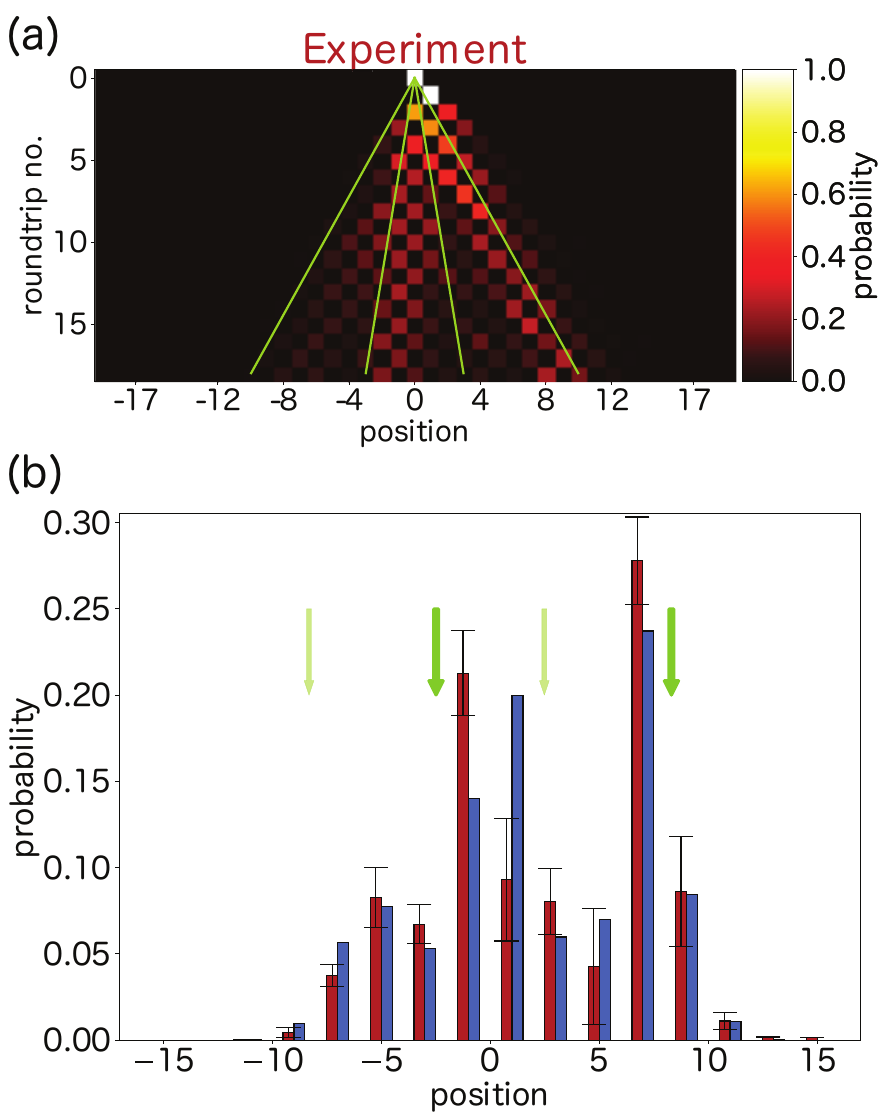}
  \caption{(a) Intensity distribution of the experimentally realized quantum walk corresponding to Fig.~\ref{fig:groupVel-numeric}, summed over all internal degrees of freedom. The indicated solid lines correspond to the asymptotic peak trajectories and are displayed to guide the eye. The average similarity to the corresponding portion of the numerical results is $\bar{\mathcal{S}}=80.1\,\%$. (b) Bar chart presentation of the position distribution at step number 15 extracted from the numerical data of Fig.~\ref{fig:groupVel-numeric} (blue) and the experimental data of Fig.~\ref{fig:groupVel}a (red). The arrows indicate the positions and the expected relative intensities expected from the asymmetric input state of the walk. The observed central peak is a result of the overlap and interference of the two slower wavefronts, marking a clear distinction from the double-lobed distribution of any standard or split-step quantum walk. Error bars were obtained by a Monte Carlo simulation of the effects of uncertainties of $\pm 1^{\circ}$ in the three coin angles, and uncertainties of $\pm 2.5\,\%$ of the four detection efficiencies (see \ref{sec:errordisc}).}
	\label{fig:groupVel}
\end{figure}

In this section we report on a dynamical feature genuine to four-dimensional coins.
As pointed out in Sec.~\ref{sec:theo_gv}, such coins can give rise to multiple wavefronts in the position distribution of the quantum walker, i.e.\ after tracing out for the coin degrees of freedom.
Since dispersion analysis providing the wavefront structure and dynamics is accurate only in the long-time limit, the DTQW parameters has to be chosen carefully to allow sufficient resolution of the peaks within the experimental time-scale.
Our strategy is to tune the coin parameters such that the we obtain the largest difference in propagation speeds.

We consider the four-dimensional coin operator implemented by placing a HWP in the loop, corresponding to $C_L = \Chwp(20^\circ)$, and two quarter-wave plates in each arm, aligned at $27^\circ$ and $0^\circ$, respectively, corresponding to $C_A = C_B= \Cqwp(27^\circ) \Cqwp^2(0^\circ) \Cqwp(27^\circ)$ .
The dispersion spectrum of the DTQW with these parameters is depicted in Fig.~\ref{fig:dispersion-4D}b, where we can clearly resolve the effect of level repulsions.
With this choice we have reduced the number of distinct propagation velocities from eight to four ($\pm0.1655$ and $\pm0.5538$), due to degeneracies.
While experimentally only 18 steps are reachable, we have numerically calculated evolution of the intensities for 50 steps to compare to the results of the asymptotic analysis.
The results of this calculation are presented on Fig.~\ref{fig:groupVel-numeric}, with solid green lines indicating the peak positions given by the asymptotic analysis.
While the two faster peaks separate quickly within the experimentally achievable domain (indicated by a horizontal dashed line), the two slower peaks separate only after about 35 steps.
Therefore, we can expect to be able to resolve three peaks in the experimental data: the two outer peaks corresponding to the faster wavefronts, and a single peak in the middle resulting from the transient overlap and interference between the two slower wavefronts.

The experimental results for the complete evolution are depicted in Fig.~\ref{fig:groupVel}a.
We can observe that the propagation velocities of the two outer peaks closely match the asymptotically expected values.
To offer a direct comparison of the numerical and experimental data we present the respective probability distributions after 15 steps as a bar chart plot in Fig.~\ref{fig:groupVel}b.
In addition to the numerical distribution, we have indicated the peak positions yielded by the asymptotic analysis by vertical arrows.
The positions and intensities of the outer peaks appear to be robust to the unavoidable experimental imperfections (affecting both the evolution and the initial state).
The central peak structure shows greater sensitivity: while the numerical results display a more dominant right wavefront, in the experiment the left propagating one appears to be dominating.

While the overlap and interference prevent the resolution of the positions and intensities of the two inner peaks, the presence of more than two wavefronts proves the realization of a DTQW with a genuine four-dimensional coin.

\subsection{Quantum walks on circles}
\label{sec:CircleWalks}
\begin{figure}
	\includegraphics[width=.9\columnwidth]{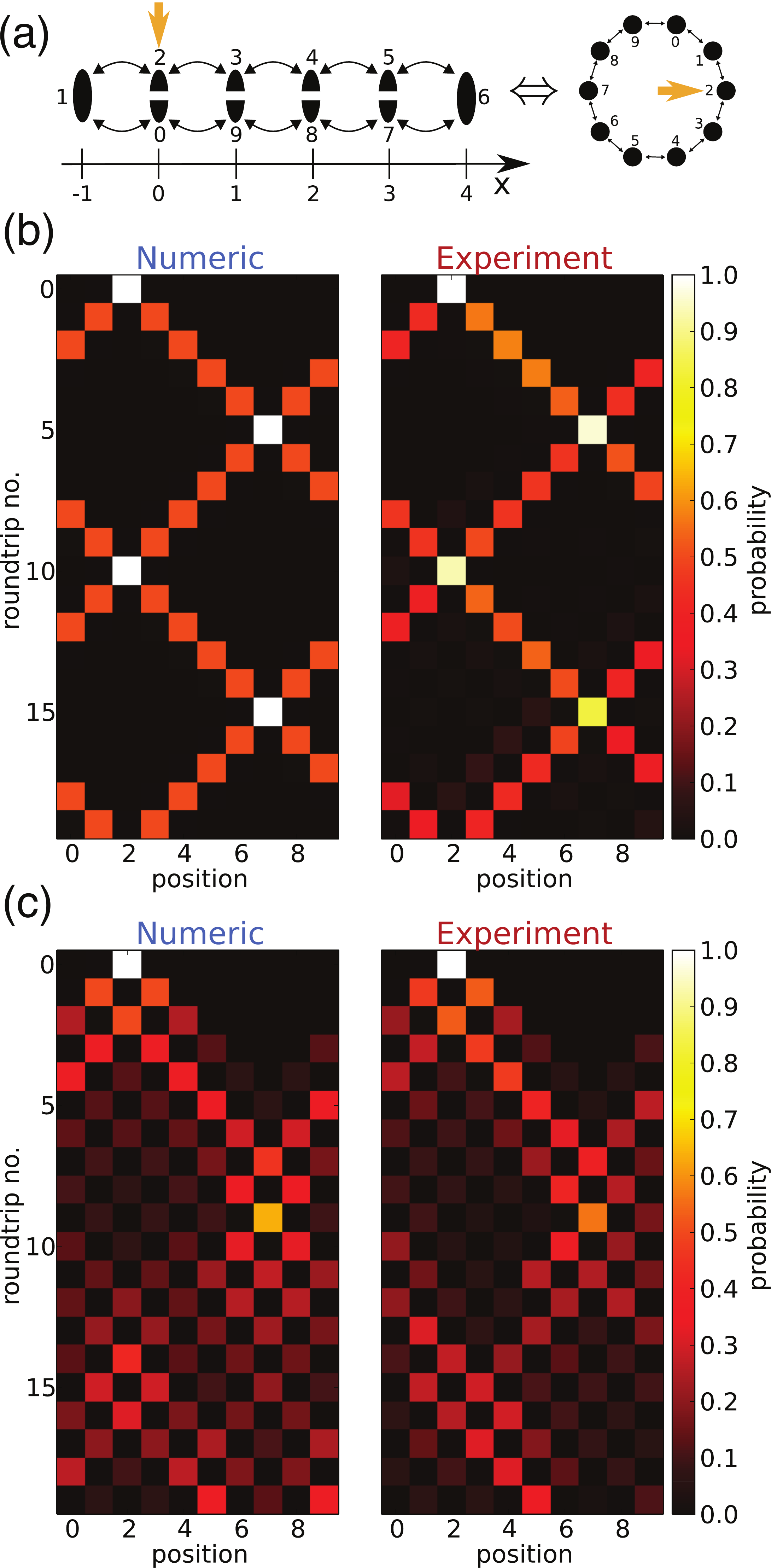}
	\caption{(a) Configuration of an asymmetric circle walk on 10 sites. The walk is started at $x=0$ (orange arrow), which corresponds to position $m = 2$ on the circle and in the chessboard plot axes. The input state is $\ket{\ccV}$ and (b) shows the step evolution of a non-mixing walk (similarity averaged over 19 steps: $\bar{\mathcal{S}}=88.9\,\%$), while (c) displays the evolution of an effective Hadamard walk in such a configuration (similarity: $\bar{\mathcal{S}}=79.9\,\%$).}
	\label{fig:Circ_Asym}
\end{figure}
\begin{figure}
\includegraphics[width=.9\columnwidth]{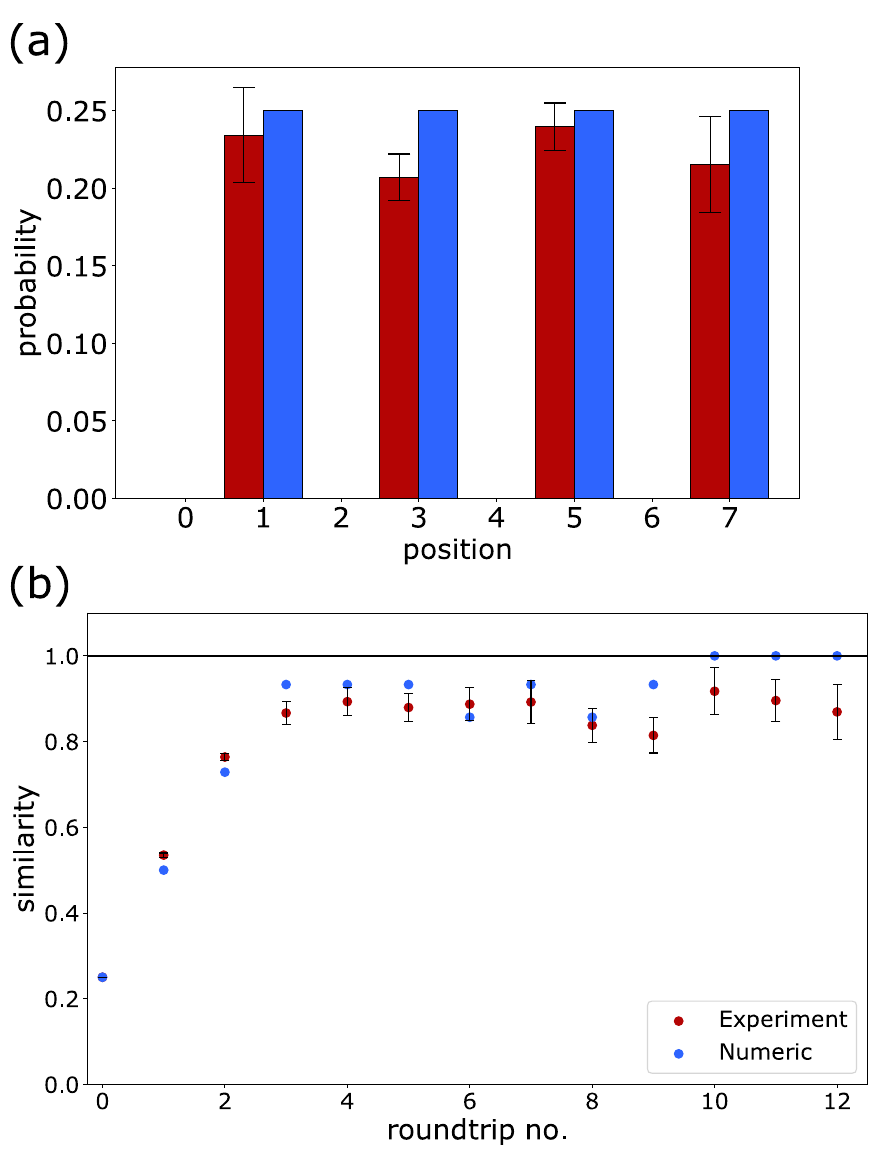}
	\caption{Equidistribution for a QW on an 8-site. Upper Panel: Intensity distribution in step 11 of the experimental (red) and numerical (blue) data for almost ideal mixing (polarization is traced out).
		Lower panel: Similarity to the flat distribution on the relevant positions $1,3,5,7$ of the experimental (red dots) and numerical (blue dots) data, plotted versus the roundtrip number.
		In both cases the deviation of the experimental data from the numeric can be explained through imperfect switchings at the boundaries such that a small part of the intensity leaves the circle sites. Since we present the data without renormalization over the circle sites only, but take the ``lost'' intensity into account, the walker's overall intensity over the circle positions only is less than 1. For original data and description of the error bars see Fig.~\ref{fig:cycleQW8sites} and \ref{sec:errordisc}.}
	\label{fig:mixing}
\end{figure}

With the large degree of coherence provided by the Michelson loop for static coins confirmed, we focus on harnessing the possibilities offered by dynamically control of all three EOMs shown on Fig.~\ref{fig:Setup}.
We have developed a scalable technique to use the additional control and the higher dimensional coin space to efficiently realize DTQWs on cyclic graphs of various topologies.

A circle graph, while locally appearing as one dimensional, requires the 2D Euclidean space to be embedded into.
However, DTQWs on circles can still be implemented using well-chosen dynamics on a 1D line graph, either by exploiting the bipartite structure \cite{bian_experimental_2017}, or as we explain below, using additional coin degrees of freedom.
Separating the two halves of the coin space based on the propagation directions enables the implementation of DTQWs on two parallel 1D lines (see Sec.~\ref{sec:HadWalks}).
By pairwise connecting the ends of these lines at appropriately chosen positions with the help of controlled operations the 1D positions can be mapped to the upper and lower arcs of a circle.
The general and scalable technique can be used to implement DTQW dynamics on circles with position dependent coin operators and arbitrary sizes.
In the experiments reported here, the DTQW dynamics on circles satisfy periodic boundary conditions.
However, with the mapping to an underlying periodic spatial structure more general, twisted boundary conditions can be realized as well \cite{alberti_quantum_2017}.

The technique to realize circles involves position dependent coins which we have implemented by three fast-switching EOMs, each realizing a specific polarization rotation according to Eq.~\eqref{eq:EOM}.
EOMs are placed in the loop and each of the two arms, along with the static waveplates.
In Fig.~\ref{fig:Circ_Asym}a we demonstrate how a circle can be formed in the graph of Fig.~\ref{fig:schemes} (d) by choosing two end points (here: $x = -1$ and $x = 4$), allowing no coupling between \cw\ and \cc\ components in the inner positions and no coupling from the end positions outwards.
This leaves an effective walk on a circle of $2N$ sites if the two endpoints are $N$ positions apart.
In the following we label the sites using a coordinate $m = 0$ through $m = 2N-1$.
We can describe this walk using a two-dimensional coin and a step operator as in Eq.~\eqref{eq:effCS}, but with an additional periodic boundary condition $\ket{m} \equiv \ket{m + 2N}$.

We have measured the results of applying both mixing and non-mixing operations on the circle.
Note that instead of the conventional Hadamard operation as given in Eq.~\eqref{eq:effCS} we here use another balanced matrix with different complex phases,
\begin{equation} \label{eq:Hadlike}
H^\prime=\Ceom(45^\circ) = \frac{1}{\sqrt{2}}\begin{pmatrix} 1 & -i \\ -i & 1
\end{pmatrix}.
\end{equation}
This is because the coin matrix in Eq.~\eqref{eq:effCS} cannot  be directly realized by an EOM, which we need for the position dependence. Note that this gives the same 50:50 splitting and as such we refer to \eqref{eq:Hadlike} as Hadamard-like coin.
For the different settings and the associated physical implementation see table~\ref{tab:CoinSettings}.
\begin{table}
\begin{flushleft}
(a) \textbf{Non-mixing}
\end{flushleft}	
	\begin{tabular}{| p{2.3cm} | p{3.1cm} | p{2.5cm} |}
\hline
 & arms & loop \\
\hline \hline
\multicolumn{3}{|l|}{\emph{inner positions}}  \\
\hline
static elements & $\left[ \Cqwp (45^\circ)\right]^2$ & $\Chwp (45^\circ) $ \\ 
EOM & $\Ceom (0^\circ)$ & $\Ceom (0^\circ) $ \\ 
resulting action & $C_A=C_B = -iX$ & $C_L = X$ \\
\hline\hline
\multicolumn{3}{|l|}{\emph{end positions}}  \\
\hline
static elements & $\left[ \Cqwp (45^\circ)\right]^2$ &  $\Chwp (45^\circ) $ \\
EOM & $\Ceom (-90^\circ) $ & $\Ceom (-90^\circ) $ \\ 
resulting action & $C_A=C_B = \mathds{1}$ & $C_L = i\mathds{1}$ \\ 
\hline 
\end{tabular}
\begin{flushleft}
  (b) \textbf{Hadamard-like}
\end{flushleft}	
\begin{tabular}{| p{2.3cm} | p{3.1cm} | p{2.5cm} |}
\hline 	
  & arms & loop  \\
\hline \hline 
\multicolumn{3}{|l|}{\emph{inner positions}}  \\
\hline
static elements &$\left[ \Cqwp (45^\circ)\right]^2$ & $\Cqwp (45^\circ)$ \\
EOM & $\Ceom (0^\circ) $ & $\Ceom (0^\circ) $ \\
resulting action & $C_A=C_B =  -iX$  & $C_L = H^\prime$\\
\hline \hline 
\multicolumn{3}{|l|}{\emph{end positions}}  \\
\hline
static elements &$\left[ \Cqwp (45^\circ)\right]^2$ & $\Cqwp (45^\circ)$ \\
EOM & $\Ceom (-45^\circ) $ & $\Ceom (-45^\circ) $ \\
resulting action & $C_A=C_B = H^\prime$ & $ C_L = \mathds{1}$   \\ 
 
\hline 
\end{tabular}
\caption{Experimental realization of the coin settings for QWs on circles for non-mixing and Hadamard-like operation. The static elements act the same way in every position, while the dynamic EOM can perform distinct operations for the inner and end positions. The total action can be computed by taking the products of static WP and EOM matrices given in Eqs.~\eqref{eq:QWP}--\eqref{eq:EOM}. Each WP needs to be considered once in the loop, and twice in the arms due to the reflection, giving rise to the square of the operators. Note that the EOM is only switched on for one direction (when the pulses pass it after the reflection at the mirror) in order to keep the number of overall switches low. The resulting operations are indicated with identity $\mathds{1}$, the Pauli $X$ gate and $H'$ as given in Eq.~\eqref{eq:Hadlike}.}
\label{tab:CoinSettings}
\end{table}
In Fig.~\ref{fig:Circ_Asym} we plot the intensity evolution of the walk on an 10-node circle for both the non-mixing and the $H'$ operation, for which we need to employ all three EOMs for the dynamic switchings, discriminating between inner and boundary positions of the graph. 
We plot only over the relevant positions $m = 0,\dots,9$ and find a high agreement between experiment and numerics.

A characteristic effect observable in walks on certain circle graphs is the so-called \textit{equidistribution} or equalization, meaning that the probability distribution corresponding to the wave function becomes close to uniform.
In an earlier work a similar effect has been studied in QWs on the line \cite{nayak_quantum_2000}, where the term \textit{mixing} was used.
We, however, find it more appropriate to reserve the use of the term mixing for a property that arises as a time average \cite{aharonov_quantum_2001, tregenna_controlling_2003, bednarska_quantum_2003, tregenna_controlling_2003}, acknowledging that unitary processes generally do not converge to a stationary distribution.
We analyze the equidistribution in detail in Fig.~\ref{fig:mixing}, where we present the intensity histogram for roundtrip 11 in which the experimental data from an 8-node circle shows nearly equal intensity at all four occupied positions (see Fig.~\ref{fig:cycleQW8sites}b for the complete evolution).
We note that the equidistribution effect is not universal, and is exhibited only be circles of certain sizes, among which the 8-node circle is the largest \cite{konno_periodicity_2017}.

In the second panel we track the similarity of the walker's probability distribution (summed over the coin degrees of freedom) to the uniform distribution, as the function of the roundtrip number.
Similarity of position distributions is defined analogous to the similarity $\mathcal{S}$ defined Eq.~\eqref{eq:sim}, just with the $d$ indices dropped.
We can extract an equidistribution time of approximately 10--12 roundtrips in agreement with the numerical model.
This equidistribution effect is likely linked to the perfect state revival after 24 steps for a 8-node circle \cite{dukes_quantum_2014}: an initially localized state goes through a uniform distribution at half of the period, along with some neighboring steps. An example of a smaller circle with 4 sites showing the revival of the initial state in 8 repetitions is presented in Sec.~\ref{sec:circles}.

The technique used for implementing circle graphs is inherently scalable to realizing any circle with an even number of nodes, since the size is set by choosing the switching times of the EOMs, without needing any extra resources.
We present results for circles of sizes 8 and 16 in \ref{sec:circles}.

\subsection{Walks on figure-eight graphs}
\label{sec:EightWalks}
\begin{figure}
	\includegraphics[width=.9\columnwidth]{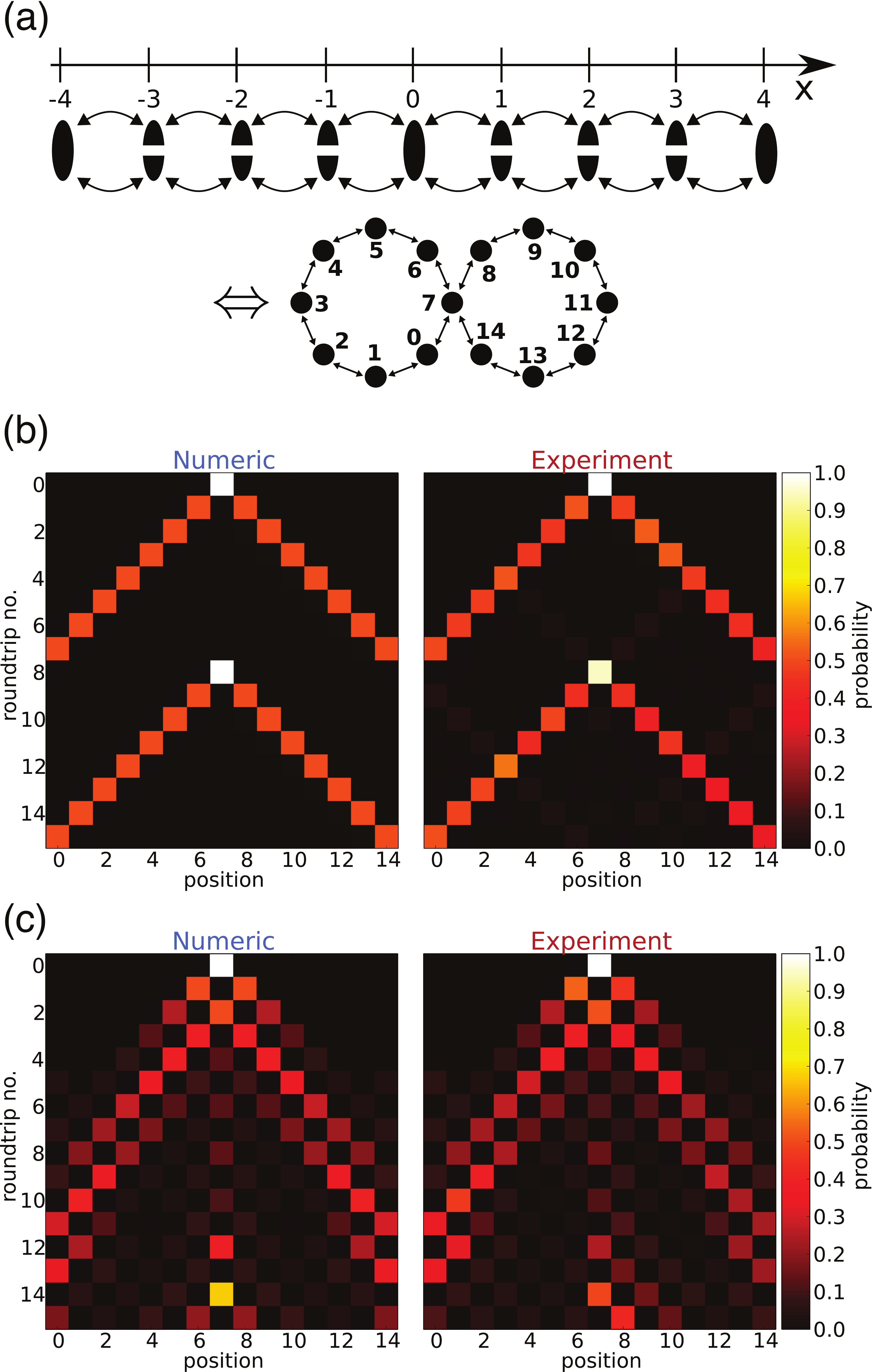}
	\caption{(a) Modified ladder graph, equivalent to a walk on a figure-eight, shown here for a graph with 15 nodes. Note the coupling at $x = 0$, which involves all four possible links. 
	(b) Numerics and experimental data of the intensity evolution of a figure-eight quantum walk with a non-mixing coin and $\ket{\ccV}$ input. The polarization resolved similarity averaged over 15 steps is $\bar{\mathcal{S}}=90.3\,\%$. 
	(c) same as (b) but with effective Hadamard splitting. (Similarity: $\bar{\mathcal{S}}=90.5\,\%$)}
	\label{fig:fig8QW}
\end{figure}
The circle graphs presented in the previous section represent a significant advance, offering a basis for simulation of systems obeying periodic boundary conditions.
Our setup, however, is capable of realizing graphs more complex than these rank-two regular graphs.
As an example, we realize DTQWs on simple instance of a Husimi cactus graph having the shape of a figure-eight, depicted on Fig.~\ref{fig:fig8QW}a, both with non-mixing and Hadamard-like coin dynamics on the circle arcs.
Due to the coupled loops, Husimi cactus graphs are studied in the context of polymer networks in solution in the field of chemical physics \cite{galiceanu_spectra_2007}, and for the interplay of search probability and centrality of the marked vertex in quantum search algorithms \cite{berry_quantum-walk-based_2010}.

To implement novel dynamics, the coin at the central rank-four node where the two circles are joined must be a genuine four-dimensional operator.
Dynamics on the nodes of the circle are experimentally implemented by dynamically controlled elements analogously to that of the circles (listed in table \ref{tab:CoinSettings}).
In order to implement the additional links at position $x = 0$ (equivalently, $m =7$) we perform in the non-mixing setting an extra switch with the EOMs in the arms by $-90^\circ$ compensating the polarization swap by the static elements $\sqrt{C_A} = \sqrt{C_B} = \Cqwp(45^\circ)$, while the loop operation is given only by the passive HWP swapping the polarization $C_L = \Chwp(45^\circ)$, thus e.g.\ the mode $|\ccH\rangle$ is mapped to $|\cV\rangle$ and vice versa.
In the Hadamard-like setting the static mixing coin in the loop $C_L = \Cqwp(45^\circ)$ is accompanied by the arm coin operations $H^\prime = \Cqwp^2(45^\circ)\cdot \Ceom(-45^\circ)$ employing again an additional EOM switch.
In the latter case, the resulting $C_A = C_B = H'$ combines with the rotation $C_L = H'$ in the loop to form a full-rank four-dimensional coin matrix (see Eq.~\eqref{eq:fullCoin})
\begin{equation}
C = \frac{1}{2} \begin{pmatrix}
1 & -i & 1 & i \\
-i & 1 & i & 1 \\
1 & i & 1 & -i \\
i & 1 & -i & 1
\end{pmatrix}.
\label{}
\end{equation}
The results for both of the settings are presented in Fig.~\ref{fig:fig8QW} (b) and (c), respectively.
One can clearly observe the light reappearing at node 7 after one cycle around the right and the left half of the figure-eight.
Again, the coherence properties ensure a high agreement of experimental and numerical data even on such a complex graph structure.
This proves the versatility of the four-dimensional coin operation compared to its two-dimensional counterpart for tailoring the ballistic spreads of a translation invariant system and the flexibility in designing non-trivial graphs.
We note that the lengths of the left and right loops could have had been chosen arbitrarily, and the present choice was made such that we can observe interference between pulses within the experimentally attainable steps.
Our work opens the route to experimental simulation of energy transport in biological structures, for example in the photosynthetic apparatus of the purple bacterium, which are modelled by coupled circular and figure-eight shaped light harvesting structures \cite{cogdell_architecture_2006, baghbanzadeh_distinguishing_2016, baghbanzadeh_geometry_2016}.

\section{Conclusion and outlook}
\label{sec:discussion}

Fully exploiting the potential of discrete time quantum walks requires a reliable and comprehensive implementation of higher dimensional coin operators.
To realize DTQW dynamics with genuine four-dimensional coin operators we have developed a novel experimental platform based on a looped Michelson interferometer.
We have carried out several experiments with increasing complexity, demonstrating the accuracy, stability and capacity of our setup to realize four-dimensional coins.
We started from a conventional Hadamard walk on the line, which shows high coherence over many steps, but essentially uses only a two-dimensional subspace of the available coin operations.
Next we presented a coin implementation that exhibits multiple distinct propagation wavefronts, precluding any description as a walk with an effectively two-dimensional coin.
Based on the four-dimensional coin degree of freedom, we developed a scheme to realize quantum walks on circles with programmable sizes over many steps by dynamic coin operations, without resorting to experimentally costly multi-step schemes.
Finally, we presented a QW on an example of a Husimi cactus graph, resembling a figure eight, that involved realizing periodic boundary conditions at both ends and a central node with a coin equivalent to having links to four neighbors.
Realization of this structure required the simultaneous implementation of genuine four-dimensional and effectively two-dimensional coins during the evolution.
The flexibility of the existing setup has been demonstrated by two experiments, with dynamics on the arcs of the figure eight corresponding to a non-mixing, and a Hadamard-like coin, respectively.

The experimental platform in principle allows the realization of DTQWs with arbitrary four-dimensional coin operators, limited only by the polarization rotating elements available to the experiment.
We have proposed an explicit three-step protocol that makes use of the property that every $4\times4$ unitary can be expressed as a product of two coin operators from the class achievable in a single roundtrip.
Therefore, any coin, such as the Grover and Fourier coins are achievable, reaching far beyond the capacities of previous experiments relying on multi-step protocols based on two-dimensional coins \cite{kitagawa_observation_2012, elster_quantum_2015, bian_experimental_2017}.

Losses of the optical signal at each round trip play a critical role in the applicability of the setup.
Implementing deterministic incoupling and outcoupling instead of the partially reflecting mirror has been recently employed in the Mach--Zehnder geometry yielding nearly 40 steps \cite{nitsche_probing_2018}, by reducing the roundtrip losses below $20$\,\%.
This approach applied to the present geometry would be necessary for more advanced applications requiring long evolution times and multiple walkers.

The present experiment makes use of three EOMs each with limited switching capabilities.
If polarization components with sufficiently versatile dynamical control are available, as our theoretical results indicate above, arbitrary dynamically controlled $4\times4$ coin operators are reachable.
Such technology would enable the implementation of DTQW on a line with effective three-dimensional coins, such as lazy walks \cite{inui_one-dimensional_2005, stefanak_limit_2014, stefanak_stability_2014, dan_one-dimensional_2015}, and a simple quantum game \cite{rajendran_playing_2018}.
Appropriate switching of coins would also enable the realization of DTQW dynamics with arbitrary twisted boundary conditions \cite{alberti_quantum_2017} by implementing the underlying periodic structures, or more generally quivers \cite{derksen_quiver_2005}.

The setup could be extended to realize dynamics on 2D lattices by adding additional delay lengths, similarly to how it has been implemented in the Mach--Zehnder geometry \cite{schreiber_2d_2012, chen_observation_2018}.
The availability of arbitrary coin operators would allow for the first time the experimental study of magnetic walks  \cite{yalcinkaya_two-dimensional_2015} and the QSH topological phase of quantum walks \cite{kitagawa_exploring_2010}.
In addition, studies with four-dimensional coins enable possible observation and applications of an Anderson transition on a 2D lattice, an effect that could not be observed in split-step walks \cite{rakovszky_localization_2015}, by providing a mechanism analogous to spin-orbit coupling \cite{evangelou_anderson_1995}.
Higher dimensional coined DTQWs in 2D lattices could be combined to implement wrapped geometries allowing the experimental study of search protocols requiring periodic boundary conditions \cite{ambainis_coins_2005, hein_quantum_2010, hamilton_driven_2016}, and dynamics on M\"obius-strip like graphs \cite{li_quantum_2015}.
Additionally, implementing DTQWs on other non-trivial graph structures involving distinguished nodes with several neighbors would provide a basis for search algorithms \cite{berry_quantum-walk-based_2010} and graph isomorphism testing \cite{douglas_classical_2008}.

The above examples rely on four dimensional coin operators providing a structure to the dynamics not attainable using lower dimensional coin space dynamics.
Our platform provides the first instance of an extensible realization of a quantum walk with four-dimensional coin operators with precise dynamic control, paving the way to experimental implementations of many important applications relying on genuine higher dimensional coins.

\begin{acknowledgments}
The Integrated Quantum Optics group in Paderborn acknowledges financial support from European Commission with the ERC project QuPoPCoRN (no.\ 725366) and from the Gottfried Wilhelm Leibniz-Preis (grant number SI1115/3-1). E.~M.~S.\ acknowledges support from the Natural Sciences and Engineering Research Council of Canada (NSERC). V.~P., A.~G.\ and I.~J.\ acknowledge funding by RVO 14000; and by the project ``Centre for Advanced Applied Sciences,'' Registry No.\ CZ.02.1.01/0.0/0.0/16\_019/0000778, supported by the Operational Programme Research, Development and Education, co-financed by the European Structural and Investment Funds and the state budget of the Czech Republic. A.~G.\ and I.~J.\ have been partially supported by the Czech Science foundation (GA{\v C}R) project number 17-00844S, A.~G.\ by the National Research Development and Innovation Office of Hungary under project No.\ K124351.
\end{acknowledgments}

\appendix

\section{Theoretical details}
\label{sec:coins}

\subsection{Coins achievable in one round trip}
\label{sec:coins-one-roundtrip}

Here we prove the following related to Sec.~\ref{sec:theo}:

\textbf{Theorem 1:}
A unitary matrix
\begin{equation}
C = \begin{pmatrix}
\CA{c_{11}} & \CA{c_{12}} & \CB{c_{13}} & \CB{c_{14}} \\
\CD{c_{21}} & \CD{c_{22}} & \CC{c_{23}} & \CC{c_{24}} \\
\CD{c_{31}} & \CD{c_{32}} & \CC{c_{33}} & \CC{c_{34}} \\
\CA{c_{41}} & \CA{c_{42}} & \CB{c_{43}} & \CB{c_{44}} \\
\end{pmatrix}.
\end{equation}
can be written in the form (see Eq.~\eqref{eq:fullCoin})
\begin{equation}
C =
\begin{pmatrix}
a_{HH} & 0 & 0 & a_{HV} \\
0 & b_{VV} & b_{VH} & 0 \\
0 & b_{HV} & b_{HH} & 0 \\
a_{VH} & 0 & 0 & a_{VV} \\
\end{pmatrix}
\cdot
\begin{pmatrix}
l_{HH} & l_{HV} & 0 & 0 \\
l_{VH} & l_{VV} & 0 & 0 \\
0 & 0 & l_{HH} & l_{HV} \\
0 & 0 & l_{VH} & l_{VV} \\
\end{pmatrix},
\label{eq:thm-decomp-lab}
\end{equation}
where
\begin{equation}
\begin{pmatrix}
a_{HH} & a_{HV} \\
a_{VH} & a_{VV} \\
\end{pmatrix}, \ 
\begin{pmatrix}
b_{HH} & b_{HV} \\
b_{VH} & b_{VV} \\
\end{pmatrix}, \ 
\begin{pmatrix}
l_{HH} & l_{HV} \\
l_{VH} & l_{VV} \\
\end{pmatrix}
\end{equation}
are unitary matrices,
if and only if the matrices
\begin{equation}
\begin{pmatrix}
\CA{c_{11}} & \CA{c_{12}} \\
\CA{c_{41}} & \CA{c_{42}} \\
\CC{c_{23}} & \CC{c_{24}} \\
\CC{c_{33}} & \CC{c_{34}} \\
\end{pmatrix}, \ \begin{pmatrix}
\CB{c_{13}} & \CB{c_{14}} \\
\CB{c_{43}} & \CB{c_{44}} \\
\CD{c_{21}} & \CD{c_{22}} \\
\CD{c_{31}} & \CD{c_{32}} \\
\end{pmatrix}
\label{eq:thm-smx}
\end{equation}
both have rank one (i.e., linearly dependent rows or columns).

The implication from \eqref{eq:thm-decomp-lab} to the latter property follows trivially from performing the matrix multiplication, but it's instructive to have an explicit expansion:
\begin{equation}
\begin{aligned}
\begin{pmatrix}
c_{11} & c_{12} \\
c_{41} & c_{42} \\
c_{23} & c_{24} \\
c_{33} & c_{34} \\
\end{pmatrix} &= \begin{pmatrix}
a_{HH}l_{HH} & a_{HH}l_{HV} \\
a_{VH}l_{HH} & a_{VH}l_{HV} \\
b_{VH}l_{HH} & b_{VH}l_{HV} \\
b_{HH}l_{HH} & b_{HH}l_{HV} \\
\end{pmatrix}, \\
\begin{pmatrix}
c_{13} & c_{14} \\
c_{43} & c_{44} \\
c_{21} & c_{22} \\
c_{31} & c_{32} \\
\end{pmatrix} &= \begin{pmatrix}
a_{HV}l_{VH} & a_{HV}l_{VV} \\
a_{VV}l_{VH} & a_{VV}l_{VV} \\
b_{VV}l_{VH} & b_{VV}l_{VV} \\
b_{HV}l_{VH} & b_{HV}l_{VV} \\
\end{pmatrix}.
\end{aligned}
\label{eq:thm-cxx-lab}
\end{equation}

Let us now treat the opposite implication, i.e., assume that the two matrices in \eqref{eq:thm-smx} are of unit rank, so their elements must be of the form
\begin{equation}
\begin{aligned}
\begin{pmatrix}
c_{11} & c_{12} \\
c_{41} & c_{42} \\
c_{23} & c_{24} \\
c_{33} & c_{34} \\
\end{pmatrix} &= \begin{pmatrix}
\alpha_1 \beta_1 & \alpha_2 \beta_1 \\
\alpha_1 \beta_2 & \alpha_2 \beta_2 \\
\alpha_1 \beta_3 & \alpha_2 \beta_3 \\
\alpha_1 \beta_4 & \alpha_2 \beta_4 \\
\end{pmatrix}, \\
\begin{pmatrix}
c_{13} & c_{14} \\
c_{43} & c_{44} \\
c_{21} & c_{22} \\
c_{31} & c_{32} \\
\end{pmatrix} &= \begin{pmatrix}
\gamma_1 \delta_1 & \gamma_2 \delta_1 \\
\gamma_1 \delta_2 & \gamma_2 \delta_2 \\
\gamma_1 \delta_3 & \gamma_2 \delta_3 \\
\gamma_1 \delta_4 & \gamma_2 \delta_4 \\
\end{pmatrix}
\end{aligned}
\label{eq:thm-cxx-abcd}
\end{equation}
for some $\alpha_i, \beta_i, \gamma_i, \delta_i \in \mathbb{C}$. Any matrix formed by such elements can be written in the following product form,
\begin{equation}
C =
\begin{pmatrix}
\beta_1 & 0 & 0 & \delta_1 \\
0 & \delta_3 & \beta_3 & 0 \\
0 & \delta_4 & \beta_4 & 0 \\
\beta_2 & 0 & 0 & \delta_2 \\
\end{pmatrix} \cdot \begin{pmatrix}
\alpha_1 & \alpha_2 & 0 & 0 \\
\gamma_1 & \gamma_2 & 0 & 0 \\
0 & 0 & \alpha_1 & \alpha_2 \\
0 & 0 & \gamma_1 & \gamma_2 \\
\end{pmatrix}.
\label{eq:thm-decomp-abcd}
\end{equation}
This is already close to the desired form \eqref{eq:thm-decomp-lab} but nothing so far guarantees that the two matrices forming the right-hand side are also unitary and thus realizable by separate physical transforms.

It is easy to show that if $\alpha_1$ and $\alpha_2$, or $\gamma_1$ and $\gamma_2$, were simultaneously zero, $C$ would be singular. In all the other cases there is some freedom in decomposing the left-hand sides of \eqref{eq:thm-cxx-abcd}, so without loss of generality we can assume that $|\alpha_1|^2 + |\alpha_2|^2 = |\gamma_1|^2 + |\gamma_2|^2 = 1$.

The unitarity of $C$ postulates that the norm of its first and last row must be equal to $1$ and their scalar product must vanish. With the above assumption these equations take the forms
\begin{equation}
\begin{aligned}
|\beta_1|^2 + |\delta_1|^2 &= 1, \\ 
|\beta_2|^2 + |\delta_2|^2 &= 1, \\
\beta_1 \overline{\beta}_2 + \delta_1 \overline{\delta}_2 &= 0.
\end{aligned}
\end{equation}
This simply says that the matrix
\begin{equation}
\begin{pmatrix}
\beta_1 & \delta_1 \\
\beta_2 & \delta_2
\end{pmatrix}
\label{eq:proof1:bd1}
\end{equation}
is unitary. In \eqref{eq:thm-decomp-abcd} these four elements take positions of the elements $a_{ij}$ of \eqref{eq:thm-decomp-lab}.
Similarly from the middle two rows of \eqref{eq:thm-decomp-abcd} we derive the unitarity of
\begin{equation}
\begin{pmatrix}
\beta_3 & \delta_3 \\
\beta_4 & \delta_4
\end{pmatrix},
\label{eq:proof1:bd2}
\end{equation}
or $(b_{ij})$.

We also require that the first and second column of $C$ are normalized and orthogonal vectors.
Note that the unitarity of \eqref{eq:proof1:bd1} and \eqref{eq:proof1:bd2} also implies
\begin{equation}
|\beta_1|^2 + |\beta_2|^2 = |\beta_3|^2 + |\beta_4|^2 =
|\delta_1|^2 + |\delta_2|^2 = |\delta_3|^2 + |\delta_4|^2 = 1.
\label{eq:thm-bd-units}
\end{equation}
Using the last equality, the row orthonormality condition gives the following equations,
\begin{equation}
\begin{aligned}
|\alpha_1|^2 + |\gamma_1|^2 &= 1, \\
|\alpha_2|^2 + |\gamma_2|^2 &= 1, \\
\alpha_1 \overline{\alpha}_2 + \gamma_1 \overline{\gamma}_2 &= 0,
\end{aligned}
\end{equation}
which again are nothing else than the conditions on unitarity of
\begin{equation}
\begin{pmatrix}
\alpha_1 & \alpha_2 \\
\gamma_1 & \gamma_2
\end{pmatrix},
\end{equation}
forming the blocks of the latter matrix in \eqref{eq:thm-decomp-abcd}.

In conclusion, the conditions stated by the theorem only allow matrices exactly of the form \eqref{eq:thm-decomp-lab} where the submatrices corresponding to $C_A$, $C_B$, $C_L$ are all unitary. Their elements can easily be reconstructed using the following algorithm:
\begin{enumerate}
	\item Build the matrices \eqref{eq:thm-smx} and take \textit{any} decomposition of the form \eqref{eq:thm-cxx-abcd}, the existence of which is guaranteed by the assumptions.
	\item Multiply the vectors $(\alpha_1, \alpha_2)$ and $(\gamma_1, \gamma_2)$ by some constants to achieve norms of one, dividing $(\beta_j)$ and $(\delta_j)$ by the same constants to keep products invariant.
	\item Compare \eqref{eq:thm-cxx-abcd} with \eqref{eq:thm-cxx-lab} to find correspondence between $\alpha_j, \beta_j, \gamma_j \delta_j$ and $a_{ij}, b_{ij}, l_{ij}$.
\end{enumerate}

\textbf{Note:} The same derivation can be repeated with minimal changes when the two diagonal blocks of the latter matrix of \eqref{eq:thm-decomp-lab} are not required to be equal, only unitary (as would correspond to transforming the clockwise- and counter-clockwise-propagating pulses in the loop independently). 
The two matrices in \eqref{eq:thm-smx} then need to be replaced by four matrices
\begin{equation}
\begin{gathered}
\begin{pmatrix}
\CA{c_{11}} & \CA{c_{14}} \\ \CA{c_{21}} & \CA{c_{24}}
\end{pmatrix}, \ \begin{pmatrix}
\CB{c_{12}} & \CB{c_{13}} \\ \CB{c_{22}} & \CB{c_{23}}
\end{pmatrix}, \\
\begin{pmatrix}
\CC{c_{31}} & \CC{c_{34}} \\ \CC{c_{41}} & \CC{c_{44}}
\end{pmatrix}, \ \begin{pmatrix}
\CD{c_{32}} & \CD{c_{33}} \\ \CD{c_{42}} & \CD{c_{43}}
\end{pmatrix}.
\end{gathered}
\end{equation}

\subsection{Universality of available coins}
\label{sec:coins-universality}

The three-step protocol consists of
\begin{enumerate}[label=Step \arabic*:]
	\item apply a coin $C_1$, evolve over one round trip,
	\item let the wave packets finish one full round trip with a trivial coin,
	\item apply another coin $C_2$, finish the round trip.
\end{enumerate}
Here we rigorously prove that any $U(4)$ is experimentally attainable using this protocol.

Firstly, we point out the flip-flop nature of the step operator: two applications thereof amount to the identity map.
So if the coin is left trivial ($C_A = C_B = C_L = \mathds{1}$) in Step 2, the application of $\hat{S}$ in Step 2 negates any displacement made in Step 1 and returns the internal state to what it was immediately after the application of $\hat{C}$ in Step 1.
The state after 3 steps can be described as
\begin{equation}
\ket{\Psi_{t+3}} = \hat{S} \hat{C_2} \hat{S} \mathds{1} \hat{S} \hat{C_1} \ket{\
	Psi_t} = \hat{S} \hat{C_2} \hat{C_1} \ket{\Psi_t},
\label{eq:3step}
\end{equation}
thus the internal state is effectively transformed by the product $C_2 C_1$ and subject to just one flip-flop displacement, according to the final coin state.
The total action of these three round trips thus can be perceived as a single step of a quantum walk with a more general coin.

This coin becomes indeed completely general, if we allow the blocks of $\CLL$ (Eq.~\eqref{eq:C_LL}) to be controlled separately for the $\cw$ and $\cc$ polarizations (upper-left and lower-right blocks):

\textbf{Theorem 2:} 
Let $C$ be a generic $U(4)$ matrix. Then two transforms of the form
\begin{equation}
\begin{aligned}
C_t &=
\begin{pmatrix}
a_{HH,t} & 0 & 0 & a_{HV,t} \\
0 & b_{VV,t} & b_{VH,t} & 0 \\
0 & b_{HV,t} & b_{HH,t} & 0 \\
a_{VH,t} & 0 & 0 & a_{VV,t} \\
\end{pmatrix}
\cdot \\
&\qquad
\begin{pmatrix}
l_{HH,t} & l_{HV,t} & 0 & 0 \\
l_{VH,t} & l_{VV,t} & 0 & 0 \\
0 & 0 & l'_{HH,t} & l'_{HV,t} \\
0 & 0 & l'_{VH,t} & l'_{VV,t} \\
\end{pmatrix}
\end{aligned}
\label{eq:decomp-llab}
\end{equation}
can be found, $C_1, C_2$, such that $C = C_2 C_1$, with the individual submatrices $a, b, l, l'$ in $U(2)$. Moreover, up to a global phase correction factor, the four submatrices can be sought in $SU(2)$.

We will prove this theorem constructively, using bra-ket notation on $\mathbb{C}^2$: let in this section ket denote a two-element column vector and a bra with the same symbol its conjugate, a row vector composed of complex conjugate elements. Namely, we pair the unknowns of the decomposition in the following objects:
\begin{equation}
\begin{gathered}
\ket{p} := \begin{pmatrix} l_{HH,2} \\ l_{VH,2} \end{pmatrix}, \quad
\ket{q} := \begin{pmatrix} l_{HV,2} \\ l_{VV,2} \end{pmatrix}, \\
\ket{r} := \begin{pmatrix} l'_{HH,2} \\ l'_{VH,2} \end{pmatrix}, \quad
\ket{s} := \begin{pmatrix} l'_{HV,2} \\ l'_{VV,2} \end{pmatrix}, \\
\bra{P} := \begin{pmatrix} l_{HH,1}\ l_{HV,1} \end{pmatrix}, \quad
\bra{Q} := \begin{pmatrix} l_{VH,1}\ l_{VV,1} \end{pmatrix}, \\
\bra{R} := \begin{pmatrix} l'_{HH,1}\ l'_{HV,1} \end{pmatrix}, \quad
\bra{S} := \begin{pmatrix} l'_{VH,1}\ l'_{VV,1} \end{pmatrix}.
\end{gathered}
\end{equation}
The condition on unitarity of $\CLL$ then translates into the requirement that $(\ket{p}, \ket{q})$, $(\ket{r}, \ket{s})$, $(\ket{P}, \ket{Q})$, and $(\ket{R}, \ket{S})$ are four (not necessarily different) orthonormal bases.

We will show that the decomposition stated by Theorem 2 exists even with a further restriction
\begin{equation}
a_{ij,2} = b_{ij,2} = \delta_{ij},
\end{equation}
i.e., the $C_A$, $C_B$ matrices in Step 2 being trivial. In the following $a_{ij}$ and $b_{ij}$ will thus denote $a_{ij,1}$, $b_{ij,1}$ for brevity.

If we split the required coin matrix $C$ into $2\times2$ blocks as
\begin{equation}
C = \begin{pmatrix}
C_{TL} & C_{TR} \\
C_{BL} & C_{BR}
\end{pmatrix},
\label{eq:c-blocks}
\end{equation}
the equation
\begin{equation}
C = C_2 C_1
\end{equation}
can be expanded block-wise and written as a system of four separate block equations,
\begin{equation}
\begin{aligned}
C_{TL} &= a_{HH} \ket{p}\!\bra{P} + b_{VV} \ket{q}\!\bra{Q}, \\
C_{TR} &= a_{HV} \ket{p}\!\bra{S} + b_{VH} \ket{q}\!\bra{R}, \\
C_{BL} &= a_{VH} \ket{s}\!\bra{P} + b_{HV} \ket{r}\!\bra{Q}, \\
C_{BR} &= a_{VV} \ket{s}\!\bra{S} + b_{HH} \ket{r}\!\bra{R}.
\end{aligned}
\label{eq:block-equations}
\end{equation}

We are also given the unitarity conditions of $C$:
\begin{equation}
C^\dagger C = \mathds{1}, \quad C C^\dagger = \mathds{1}.
\end{equation}
In the block form \eqref{eq:c-blocks}, the former becomes
\begin{equation}
\begin{pmatrix}
C_{TL}^\dagger C_{TL} + C_{BL}^\dagger C_{BL} &
C_{TL}^\dagger C_{TR} + C_{BL}^\dagger C_{BR} \\
C_{TR}^\dagger C_{TL} + C_{BR}^\dagger C_{BL} &
C_{TR}^\dagger C_{TR} + C_{BR}^\dagger C_{BR}
\end{pmatrix} = \begin{pmatrix}
\mathds{1} & 0 \\
0 & \mathds{1} \\
\end{pmatrix}
\label{eq:ortho-columns}
\end{equation}
and the latter
\begin{equation}
\begin{pmatrix}
C_{TL} C_{TL}^\dagger + C_{TR} C_{TR}^\dagger &
C_{TL} C_{BL}^\dagger + C_{TR} C_{BR}^\dagger \\
C_{BL} C_{TL}^\dagger + C_{BR} C_{TR}^\dagger &
C_{BL} C_{BL}^\dagger + C_{BR} C_{BR}^\dagger
\end{pmatrix} = \begin{pmatrix}
\mathds{1} & 0 \\
0 & \mathds{1} \\
\end{pmatrix}.
\label{eq:ortho-rows}
\end{equation}

Plugging in \eqref{eq:block-equations}, we find that if such decomposition exists, it must satisfy
\begin{equation}
\begin{gathered}
|a_{HH}|^2 = |a_{VV}|^2 = 1 - |a_{HV}|^2 = 1 - |a_{VH}|^2, \\
|b_{HH}|^2 = |b_{VV}|^2 = 1 - |b_{HV}|^2 = 1 - |b_{VH}|^2.
\end{gathered}
\label{eq:complements}
\end{equation}

Along with the orthonormality of $(\ket{p}, \ket{q})$ etc., the equations \eqref{eq:block-equations} strongly resemble singular value decompositions (SVDs): indeed, they would become SVDs of the left-hand side matrices if, furthermore, the $a_{ij}$ and $b_{ij}$ coefficients were real and nonnegative. Without loss of generality, we can thus postulate that the first line is the actual SVD, i.e., that $\ket{p}$ and $\ket{q}$ are left-singular vectors, $\ket{P}$ and $\ket{Q}$ right-singular vectors and $a_{HH}$ and $b_{VV}$ the singular values of $C_{TL}$, and see if we can satisfy the other three lines with this choice.

Given $\ket{P}$ and $\ket{Q}$, we can apply both sides of the third line of \eqref{eq:block-equations} on them, obtaining
\begin{equation}
\begin{aligned}
C_{BL} \ket{P} &= a_{VH} \ket{s}, \\
C_{BL} \ket{Q} &= b_{HV} \ket{r}, \\
\end{aligned}
\end{equation}
If the magnitude of at least one of the coefficients $a_{VH}$ or $b_{HV}$ is known to be nonzero (that is, per \eqref{eq:complements}, unless the singular values of $C_{TL}$ were both $1$), the corresponding $\ket{s}$ or $\ket{r}$ is determined up to a complex phase. If both are, they are guaranteed to be orthonormal by \eqref{eq:ortho-rows} and \eqref{eq:complements}. If $a_{VH}$ or $b_{VH}$ is zero, we complement $\ket{s}$ as an orthonormal partner of $\ket{r}$ or vice versa, respectively, with an arbitrary phase.
In either case, the choice of the phase of the two vectors leaves $a_{VH}$ and $b_{HV}$ completely determined.
The case $a_{VH} = b_{HV} = 0$ will be handled separately near the end of the proof.

Taking the Hermitian conjugate of the second equation of \eqref{eq:block-equations} we find the vectors $\ket{r}$ and $\ket{s}$ and the numbers $a_{HV}$, $b_{VH}$ in a complete analogy to the above, leaving the same exceptional case.

After these steps, the last equation does not contain any undetermined vectors, so we need to prove that it is not a contradiction. 

Assume $a_{HH} < 1$. Then both $a_{VH}$ and $a_{HV}$ are nonzero and $\ket{s}$ and $\ket{S}$ satisfy
\begin{equation}
\ket{s} = \frac{1}{a_{VH}} C_{BL} \ket{P}, \quad
\ket{S} = \frac{1}{a_{HV}^\ast} C_{TR}^\dagger \ket{p}.
\label{eq:Ss-from-Pp}
\end{equation}
We can then study
\begin{equation}
\begin{aligned}
C_{BR} \ket{S}
&= \frac{1}{a_{HV}^\ast} C_{BR} C_{TR}^\dagger \ket{p}
= -\frac{1}{a_{HV}^\ast} C_{BL} C_{TL}^\dagger \ket{p} \\
&= -\frac{a_{HH}^\ast}{a_{HV}^\ast} C_{BL} \ket{P}
= -\frac{a_{HH}^\ast a_{VH}}{a_{HV}^\ast} \ket{s},
\end{aligned}
\end{equation}
where from step to step we used \eqref{eq:Ss-from-Pp}, \eqref{eq:ortho-rows} (lower-left block), \eqref{eq:block-equations} (first line), and \eqref{eq:block-equations} (third line). This shows that $C_{BR}$ indeed maps $\ket{S}$ to a multiple of $\ket{s}$, as \eqref{eq:block-equations} requires, but also gives a concrete value to $a_{VV}$ and shows, along with \eqref{eq:complements}, that $a_{ij}$ together form a $U(2)$ matrix.

If $a_{HH} = 1$ (but $b_{VV} < 1$), we can't use \eqref{eq:Ss-from-Pp}, but we have
\begin{equation}
\ket{r} = \frac{1}{b_{HV}} C_{BL} \ket{Q}, \quad
\ket{R} = \frac{1}{b_{VH}^\ast} C_{TR}^\dagger \ket{q}.
\label{eq:Ps-from-pS}
\end{equation}
We can still prove that $C_{BR}$ acting on $\ket{S}$ produces \textit{some} vector orthogonal to $\ket{r}$, which in turn is a multiple of $\ket{s}$: for this, we consider
\begin{equation}
\begin{aligned}
\braket{r|C_{BR}|S} &= \frac{1}{b_{HV}^\ast} \braket{Q|C_{BL}^\dagger C_{BR}|S} = \\
&= -\frac{1}{b_{HV}^\ast} \braket{Q|C_{TL}^\dagger C_{TR}|S} = 0.
\end{aligned}
\end{equation}
The last step follows from the fact that for $a_{HH}$ equal to 1, $a_{HV} = 0$ and thus $C_{TR}\ket{S} = 0$. We don't learn the phase of $a_{VV}$, as it depends on the arbitrary phases of both $\ket{S}$ and $\ket{s}$, but with $a_{HV} = a_{VH} = 0$ the $a_{ij}$ matrix is unitary for any choice.

The emergence of the last term of the last equation of \eqref{eq:block-equations} and the value of $b_{HH}$ are handled similarly, resulting in $b_{ij}$ being unitary and the system \eqref{eq:block-equations} being consistent with our solution. Since the properties of $\ket{p}, \ket{q}, \ldots, \ket{R}, \ket{S}$ also guarantee unitarity of $l_{ij,1}$, $l'_{ij,1}$, $l_{ij,2}$, and $l'_{ij,2}$, this completes the decomposition.

We left out only one special case, $a_{HH} = b_{VV} = 1$. But this case is trivial: now $C_{TL}$ is of the form
\begin{equation}
C_{TL} = \ket{p}\!\bra{P} + \ket{q}\!\bra{Q}
\end{equation}
and so is unitary, $C_{TR}$ and $C_{BL}$ are zero, and $C_{BR}$ is unitary again. Matrices of this block form can be realized in a single round trip; if necessary, a three-step protocol can be made trivially by taking $C_1 = C$, $C_2 = \mathds{1}$. This last remaining case finishes the main part of the proof.

Restricting the $a$, $b$, $l$, $l'$ submatrices to be special unitary is easy by the degrees of freedom encountered throughout the construction above. We will first consider the generic case where the off-diagonal blocks of $C$ are nonzero.

We can follow closely the same algorithm as above but in the beginning, instead of using the SVD of $C_{TL}$ directly, we fix the phases of the basis vectors so that the matrices $l_{ij,2} = (\ket{p} \ket{q})$ and $l'_{ij,1} = (\ket{R} \ket{S})^\dagger$ become unimodular. This in general changes the complex phases of $a_{VH}$ and of $b_{HV}$. In the next steps we also choose the new base pairs so that they form matrices of determinant 1. 

This only leaves the choice of balancing the phase between the two vectors in each of the four pairs. For example, multiplying $\ket{p}$ by $e^{i\varphi}$ and $\ket{q}$ by $e^{-i\varphi}$ leaves $(l_{ij,2})$ unimodular and becomes a no-operation if compensated by simultaneously multiplying $a_{HH}$ and $a_{HV}$ by $e^{-i\varphi}$ and $b_{VH}$ and $b_{VV}$ by $e^{i\varphi}$. But this amounts to a phase change in one row of the matrix $a_{ij}$ and the opposite phase change in one row of $b_{ij}$. In such a transform, all the matrices keep their determinants except the latter two, whose determinants are modified by mutually opposite phases. At a certain phase the determinants become equal, and the common phase of the two matrices can be factored out of the decomposition as a unphysical complex prefactor.

In the special case
\begin{equation}
C = \begin{pmatrix}
C_{TL} & 0 \\ 0 & C_{BR},
\end{pmatrix}
\end{equation}
we find angles $\alpha$, $\beta$ such that
\begin{equation}
\begin{aligned}
\det C_{TL} &= e^{2i(\alpha + \beta)}, \\
\det C_{BR} &= e^{2i(\alpha - \beta)}.
\end{aligned}
\end{equation}
Then
\begin{equation}
C = e^{i\alpha}
\diag\{e^{i\beta}, e^{i\beta}, e^{-i\beta}, e^{-i\beta}\}
\begin{pmatrix}
e^{-i\beta} C_{TL} & 0 \\
0 & e^{i\beta} C_{BR}
\end{pmatrix},
\end{equation}
which corresponds to choosing
\begin{equation}
\begin{aligned}
\begin{pmatrix} l_{HH,1} & l_{HV,1} \\ l_{VH,1} & l_{VV,1} \end{pmatrix} &= e^{-i\alpha} C_{TL}, \\
\begin{pmatrix} l'_{HH,1} & l'_{HV,1} \\ l'_{VH,1} & l'_{VV,1} \end{pmatrix} &= e^{-i\beta} C_{BR}, \\
\begin{pmatrix} a_{HH} & a_{HV} \\ a_{VH} & a_{VV} \end{pmatrix} &= \diag\{ e^{i\beta}, e^{-i\beta} \}, \\
\begin{pmatrix} b_{HH} & b_{HV} \\ b_{VH} & b_{VV} \end{pmatrix} &= \diag\{ e^{-i\beta}, e^{i\beta} \}, \\
\begin{pmatrix} l_{HH,2} & l_{HV,2} \\ l_{VH,2} & l_{VV,2} \end{pmatrix} &= \mathds{1}, \\
\begin{pmatrix} l'_{HH,2} & l'_{HV,2} \\ l'_{VH,2} & l'_{VV,2} \end{pmatrix} &= \mathds{1}, \\
\end{aligned}
\end{equation}
all of which are unimodular, as required.

\subsection{Split-step walks feature only two wavefronts}
\label{sec:ss-disp}

In this section we show that split-step walks are limited to exhibit two counter-propagating wavefronts.
The analysis is based on the dispersion relation of the quantum walk operator $U$.
The relation can be obtained for translation invariant DTQW considering the Fourier transform $U(k)$ of the walk operator, and calculating the eigenvalue spectrum $\lambda_j(k) = e^{i\omega_j(k)}$ for each $k \in [-\pi, \pi)$.
The group velocity defined as $v_g(k) = \omega'(k) = d\omega(k)/dk$ plays an important role in determining the propagation speeds of the wavefronts \cite{ambainis_one-dimensional_2001}.
The wavefront velocities are given by the set $\{v_g(k) \mathop{|}  k \in [-\pi, \pi) \textrm{ s.t. }\omega''(k) = 0 \}$, therefore, the number of distinct wavefronts strongly depends on the number of solutions to the equation 
\begin{equation}
\omega''(k) =0.
\label{eq:inflection}
\end{equation}

Split-step walks are defined by a walk operator of the form
\begin{equation}
U = S_+ C_2 S_- C_1,
\label{eq:U-ss}
\end{equation}
where $S_+$ and $S_-$ respectively shift the $\ket{R}$ component to the right, and the $\ket{L}$ component to the left, while leaving the other component unchanged.
The two coin operators can be taken to be $SU(2)$ matrices, thus described by pairs of complex parameters satisfying $|u_1|^2+|v_1|^2=|u_2|^2+|v_2|^2=1$, as usual.
The eigenvalues of the operator in Eq.~(\ref{eq:U-ss}) obey the equation
\begin{equation}
\cos \omega_{\pm}(k) = \pm \cos(k+\varphi) \tilde{u} + \tilde{v},
\end{equation}
where $\varphi = \arg (u_1u_2)$, $\tilde{u} = |u_1 u_2|$ and $\tilde{v} = \Re(v_1 v_2^*)$.
Formally solving Eq.~(\ref{eq:inflection}) yields
\begin{equation}
\cos(k+\varphi) = a \pm \sqrt{a^2-1},
\end{equation}
with $a=(\tilde{u}^2+ \tilde{v}^2 - 1)/2 \tilde{u}\tilde{v}$ being a real number.
To obtain a real solution for $k$, the right hand side of the equation must be real.
This is fulfilled only if $|a|\ge1$ holds.
However, the right hand side must also be between $-1$ and $1$.
Therefore, for negative $a$ the only valid equation is $\cos(k+\varphi) = a + \sqrt{a^2-1}$, while for positive $a$ it's $\cos(k+\varphi) = a - \sqrt{a^2-1}$.
This still yields two solutions for $k$ due to the cosine function being even.
However, as we shall see, this symmetry does not yield any additional wavefronts.
Indeed, the group velocities for the two bands $\omega_{\pm}$ are given by the equation
\begin{equation}
v_g^{(\pm)}(k) = \frac{ \pm \tilde{u} \sin(k+\varphi)} { 1- (\tilde{u} \cos(k+\varphi) - \tilde{v} )^2},
\end{equation}
thus while formally there are four solutions, they are pairwise degenerate since the sine function is odd.

\subsection{Wavefronts with four-dimensional coins}
\label{sec:4d-disp}

As mentioned above, calculations of DTQWs with experimentally relevant four-dimensional coin operators, given by Eq.~(\ref{eq:fullCoin}), reveals that these walks may exhibit up to eight wavefronts (i.e.\ four in each direction) in their position distribution.
Here we supply some remarks on how to interpret the band structure plots and the conclusions drawn from them.

The analysis of the dispersion spectrum, analogously to the case of split-step walks in Sec.~\ref{sec:ss-disp}, can be greatly simplified by observing that the symmetries of the evolution operator guarantee that each branch is related to each other by combinations of displacements or reflections.
Since these transformations do not affect the number of inflection points and the absolute value of the associated group velocities, a single branch can provide information about the entire structure.
Due to the vertical reflection relation, wavefronts will always appear in counter-propagating pairs, each with the same speed.

We emphasize, that the dispersion analysis merely provides the velocities of the wavefronts, and does not provide information about their characteristic widths or relative intensities, partly because these parameters depend strongly on the initial coin state.
To be able to experimentally resolve two wavefronts, the difference in their velocities must be large enough so that their relative distances exceed their characteristic widths within the experimental time frame.

\section{Additional experiments}
\label{sec:experiments}

\subsection{Partial reversal of traveling directions}
\label{sec:parrev}
\begin{figure}
	\includegraphics[width=.9\columnwidth]{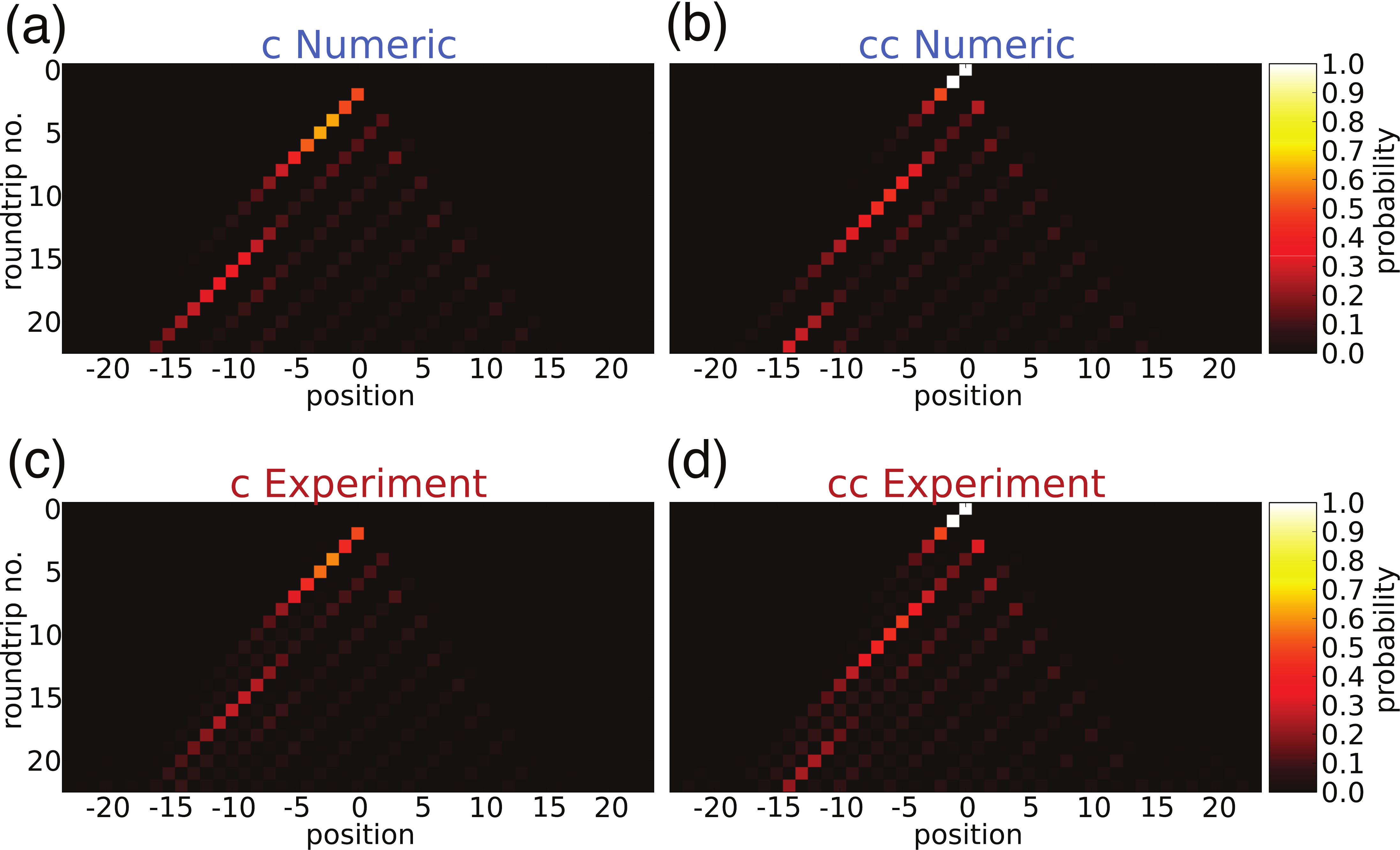}
	\caption{Numeric and experimental evolution of intensities of each traveling direction in a Hadamard walk with partial reversal of the \cw{} and \cc{} traveling directions, initialized in $\ket{\ccA}$. The sum of the two intensity distributions yields the distribution of a standard Hadamard walk. The averaged similarity between numeric and experimental results over 22 steps is $93.1\,\%$.}
	\label{fig:CplQW}
\end{figure}
Here we describe an experiment testing the coherence properties of statically coupling the traveling directions in a situation where the results have a clear intuitive interpretation.
We have set the QWP implementing operator $C_A$ to swap the polarizations, resulting in the preservation of the traveling directions as explained earlier. 
The operation $C_B$ in the other arm is implemented by a QWP at $0^\circ$, thus reversing the traveling directions in the subsequent step operation (Eq.~\eqref{eq:stepoperator}).
The situation is described by the coin operation
\begin{equation}
\CAB = \begin{pmatrix}
0 & 0 & 0 & -i \\
0 & -i & 0 & 0 \\
0 & 0 & i & 0 \\
-i & 0 & 0 & 0 \\
\end{pmatrix}.
\label{eq:C_AB_Cpl}
\end{equation}
The coin operator $\CLL$ in the loop is set to a Hadamard operation by using a HWP at $22.5^\circ$ as in Eq.~\eqref{eq:C_LL_Had}, having no effect on the travelling direction.
The obtained results are presented in Fig.~\ref{fig:CplQW}, along with numerical simulations for reference.
The strong similarity of $93.1\,\%$ between simulation and experiment proves excellent coherence properties even when both traveling directions are involved.

\subsection{Circles with 4 and 16 sites}
\label{sec:circles}

\begin{figure}
	\includegraphics[width=.9\columnwidth]{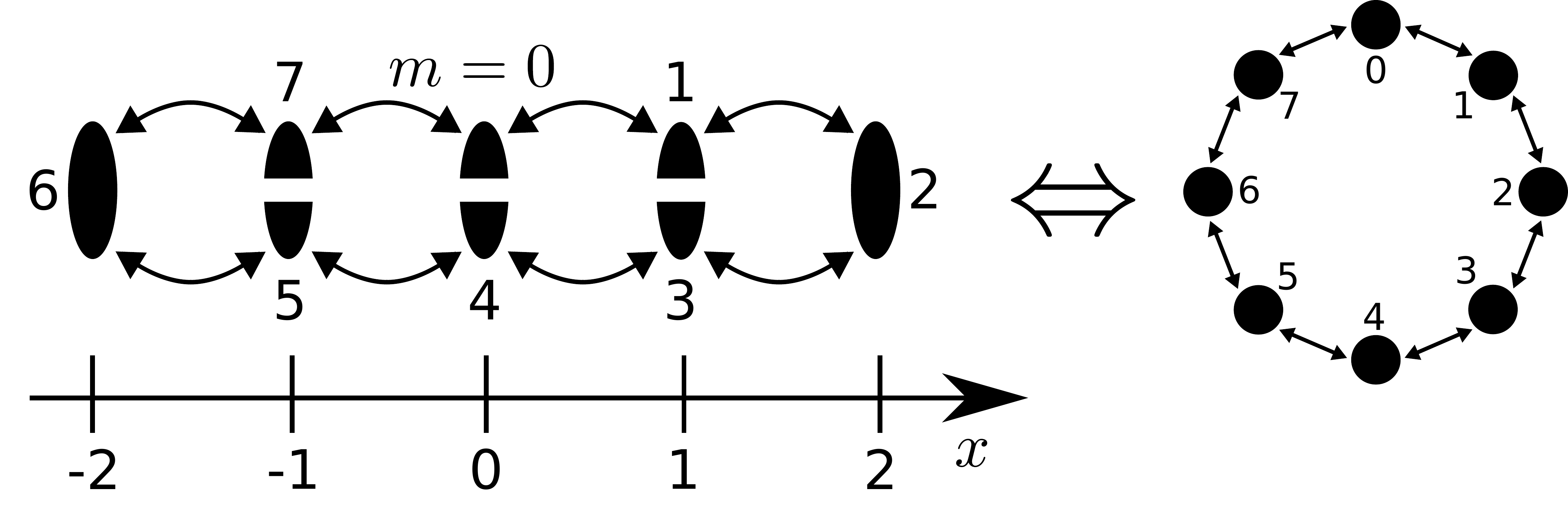}
	\caption{Modified ladder graph, equivalent to a walk on a circle, here shown for a circle with 8 nodes. The coordinate $x = -2,\ldots,2$ denotes the position on the line graph as used before while $m = 0,\ldots,7$ are the renumbered coordinates associated with the sites on the circle.}
	\label{fig:cycleQW}
\end{figure}
\begin{figure}
	\includegraphics[width=.9\columnwidth]{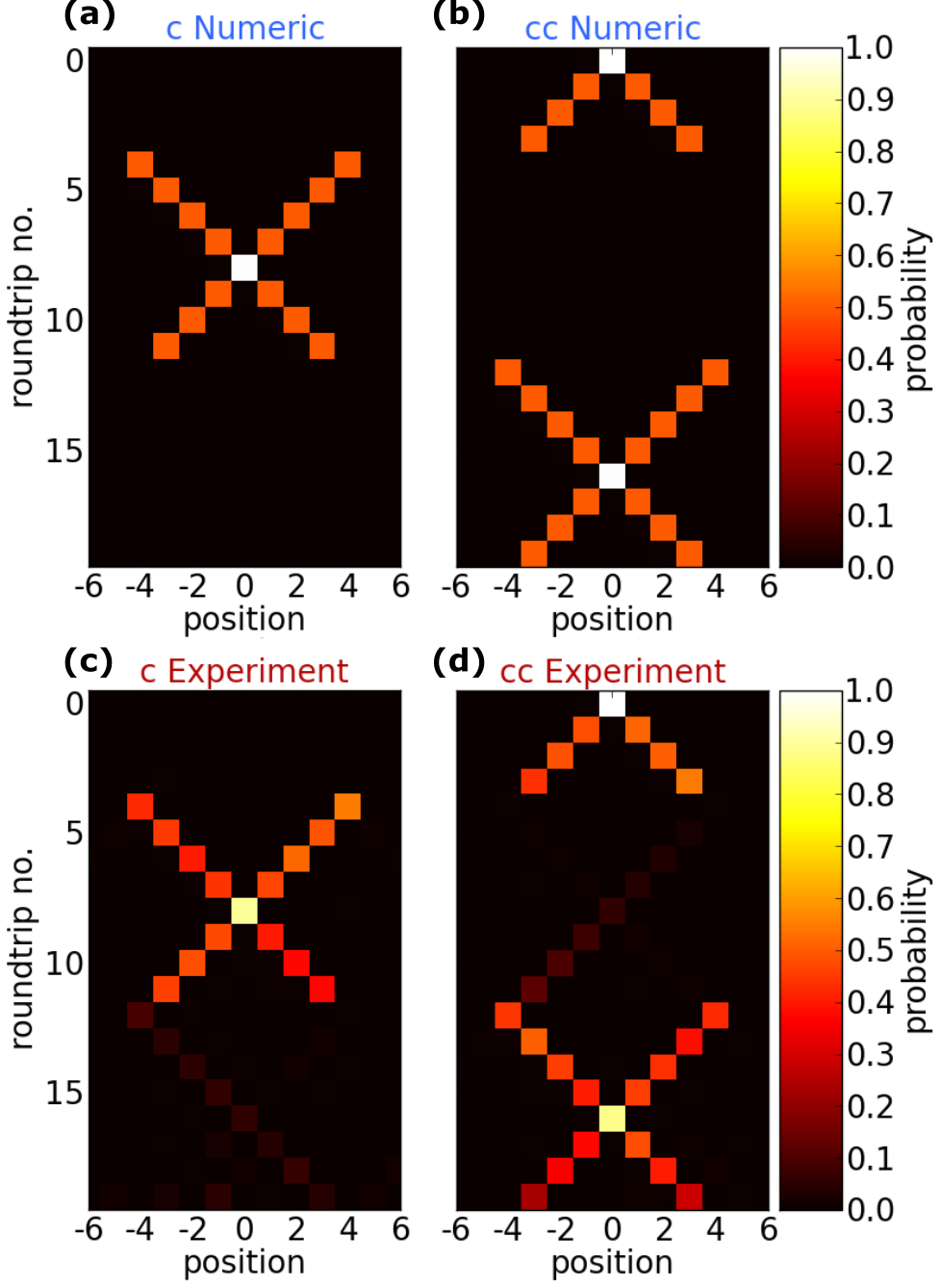}
	\caption{Walk on a circle with 16 nodes, i.e.\ with boundaries at $x = \pm4$ (non-mixing operation on the circle, $\ket{\ccV}$ input), displayed separately for clockwise and counter-clockwise propagating components in numerics and experiment. One can immediately see the jump from $\cc$ to $\cw$ (and vice versa) at $x = \pm 4$ in step 4 to 5 (step 11 to 12). Since all the outer positions are unoccupied up to small switching inaccuracies, we restrict the plot range in the other figures to the relevant positions numbered as sites on the circle $m = 0,...,15$ (as exemplified in Fig.~\ref{fig:cycleQW}). The polarization resolved similarity averaged over 19 steps is $87.1\,\%$.}
	\label{fig:cycleQW_toobig}
\end{figure}
\begin{figure}
	\includegraphics[width=.9\columnwidth]{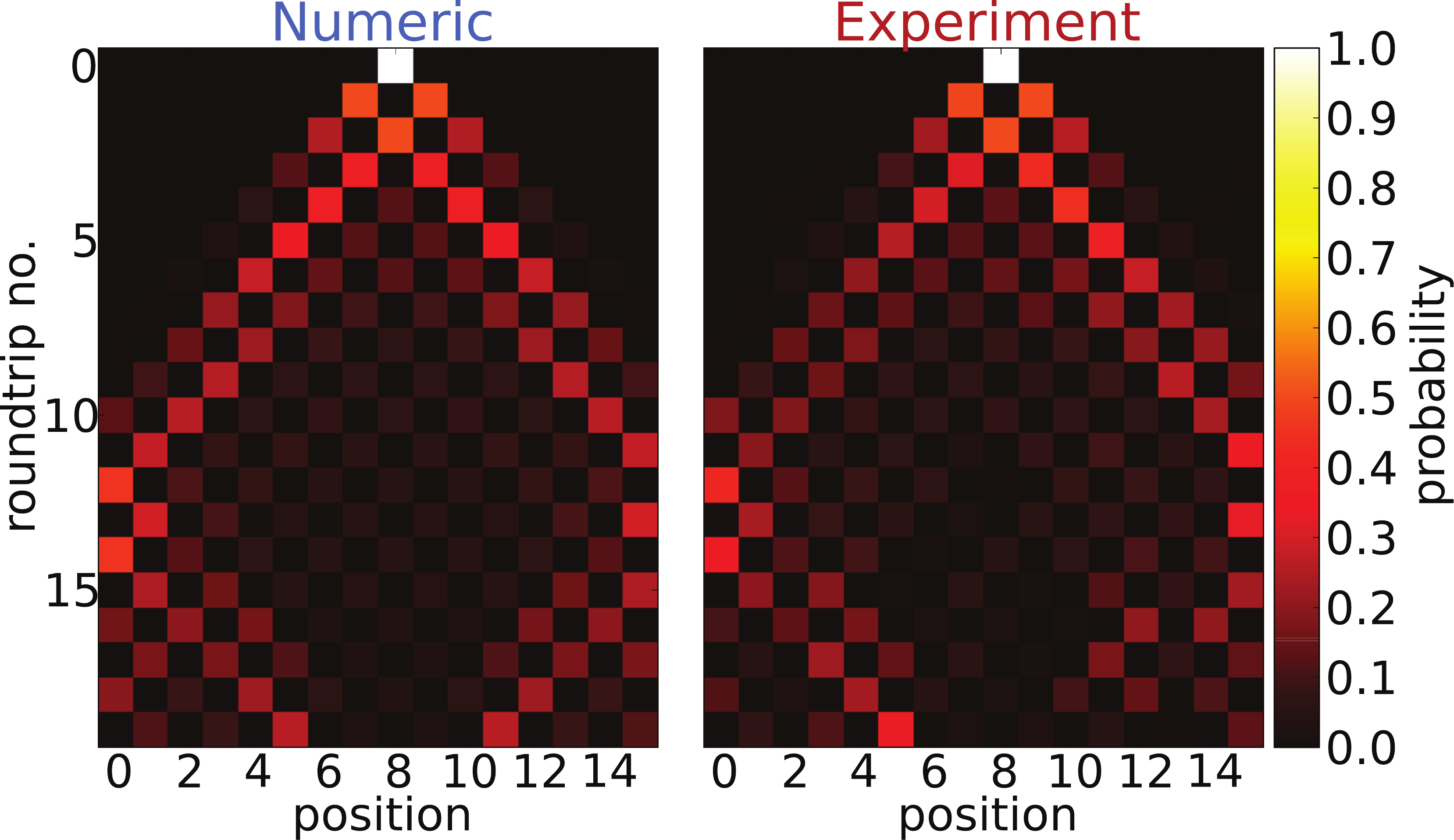}
	\caption{Hadamard walk on an 16-site circle with $\ket{\ccV}$ input. The polarization resolved similarity averaged over 19 steps is $89.1\,\%$.}
	\label{fig:cycleQW16sites}
\end{figure}
\begin{figure}
	\includegraphics[width=.9\columnwidth]{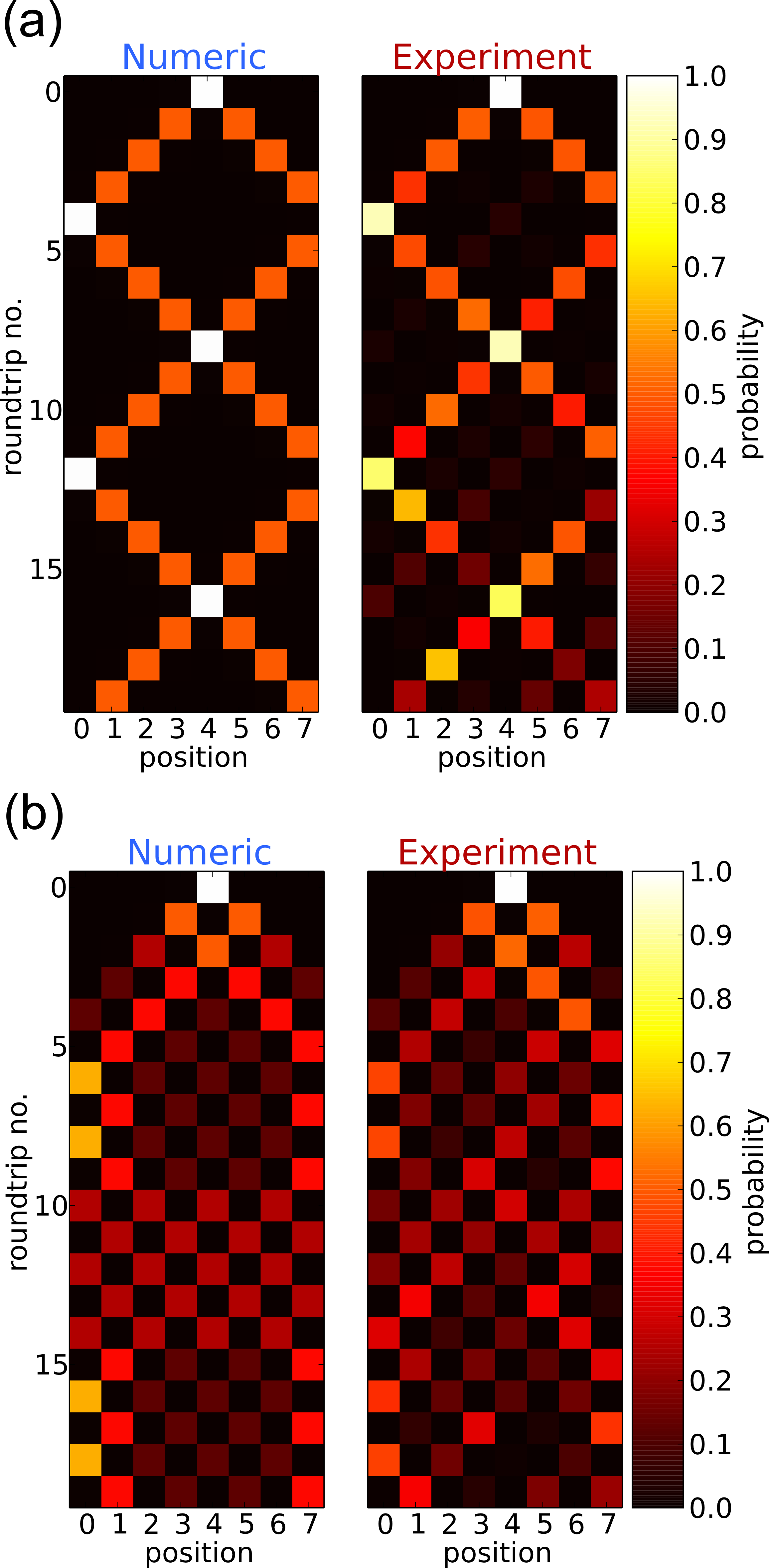}
	\caption{8-site circle (see Fig.~\ref{fig:cycleQW}): (a) Free passage walk with a similarity $\bar{\mathcal{S}} = 85.5\,\%$   and (b)  Hadamard-like $H^\prime$ walk with $\bar{\mathcal{S}} = 78.8\,\%$ (both averaged over 19 steps) with $\ket{\ccV}$ input.}
	\label{fig:cycleQW8sites}
\end{figure}
\begin{figure}
	\includegraphics[width=0.9\columnwidth]{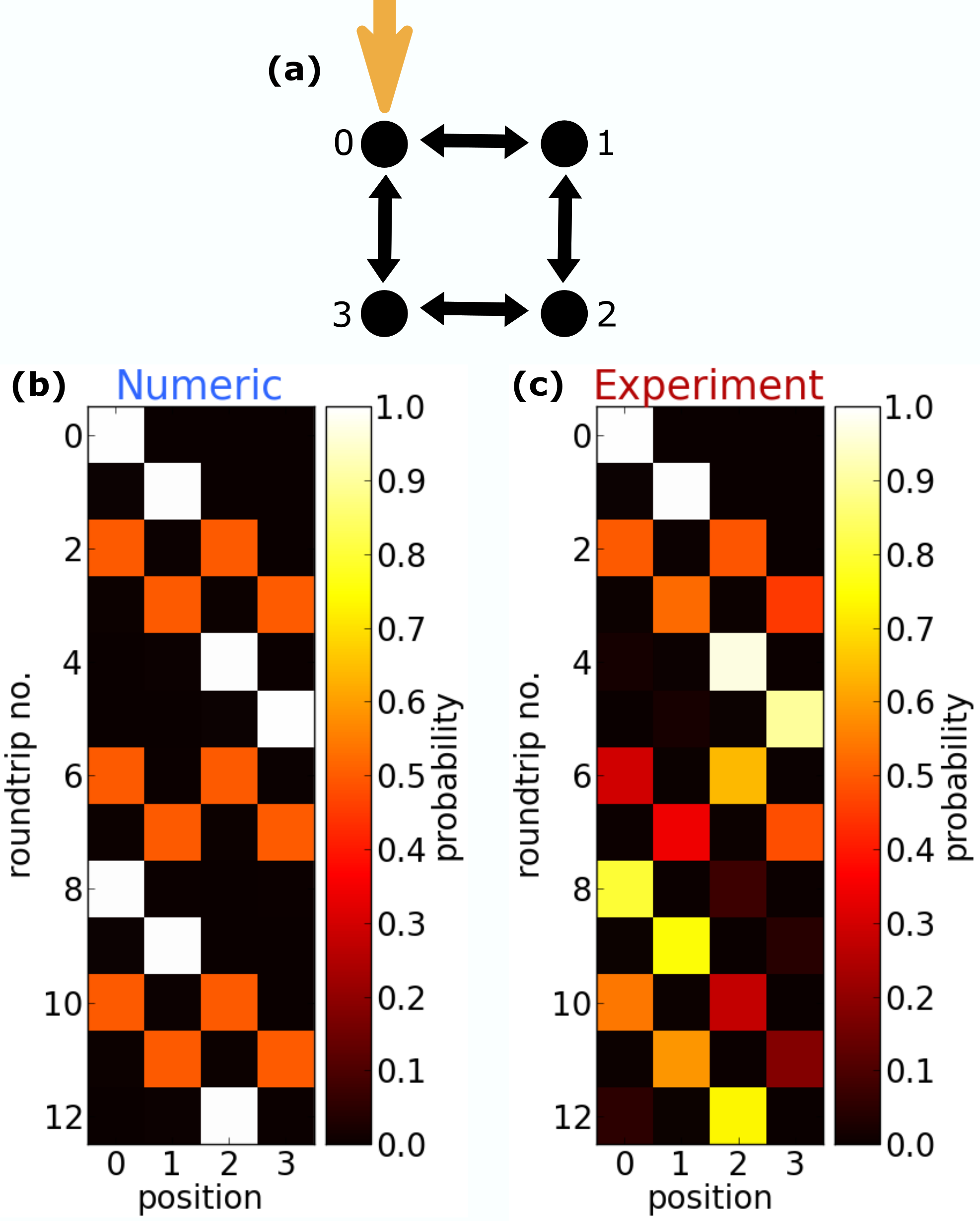}
	\caption{Hadamard walk (b) numerics and (c) experimental data of the intensity evolution for a 4-site circle (a) with  $\ket{\ccD}$ input complementing Figs~\ref{fig:Circ_Asym} and \ref{fig:fig8QW}. Polarization resolved averaged similarity over the first 12 steps is $\bar{\mathcal{S}} = 82.1\,\%$. The decrease in similarity is due to the high number of necessary EOM switches.}
	\label{fig:circleQW4sites}
\end{figure}

In this section we present data from experiments realizing walks on circles.
We recall that the inner and boundary positions are distinguished by employing all three EOMs to perform the appropriate coin operators.

In our first example we realize dynamics on a circle of size 16.
For this graph the boundaries are implemented at $x=\pm4$.
With a non-mixing coin, the input polarization $\ket{D} = \frac{1}{\sqrt{2}}(\ket{H} +\ket{V})$ in the \cc{} direction initiates two counter-propagating localized components in the walker's wave function.
In Fig.~\ref{fig:cycleQW_toobig} we present the experimental and numerical data showing how the walker is initially in the \cc{} subspace corresponding to the upper semicircle (see Fig.~\ref{fig:cycleQW} for the convention).
In the fourth step it is transferred to the lower semicircle and the two components meet again at $x=0$ corresponding to $m=8$.
The high extinction of the intensity at the ideally unoccupied positions witnesses the quality of switchings at the inner and the boundary positions. 
Results for the mixing $H'$ operation (cf. Eq.~\eqref{eq:Hadlike}) are presented in Fig.~\ref{fig:cycleQW16sites}, displaying similar features, however, with visible effects of dispersion and slower propagation speeds on the propagating waves.
Note that we applied a different plotting convention here and just display the relevant positions $m$ of the circle, see Fig.~\ref{fig:cycleQW}.

The equidistribution effect can be observed in the results presented in Fig.~\ref{fig:cycleQW8sites}c for the dynamics on a circle of 8 sites between steps 10 through 14.
Note that even (odd) positions are unreachable in odd (even) step numbers with a localized initial state, thus it is understood that the distribution is uniform over the set of positions that are allowed by the dynamics.

Implementation of a smaller circle of size 4, in which the boundary positions are at $x = 0$ and $x=1$, involves a significantly higher number of EOM switchings resulting in higher overall error, but still a similarity of more than 80\%.
The observed dynamics is presented in Fig.~\ref{fig:circleQW4sites}.
In this case an initially localized state also goes through phases of equidistribution and revival. The period of revival is 8 \cite{dukes_quantum_2014,konno_periodicity_2017}, but contrary to the previous case the equidistribution does not happen at one half of that time, but earlier at steps 2 and 3. At step 4 we instead observe a phenomenon where the probability distribution of the initial state reappears, but shifted to a node opposite the starting node. The equidistribution time, then, is one half of the first time of occurrence of this ``shifted revival''.

\subsection{Error discussion}
\label{sec:errordisc}

In this section we describe the method used for determining the extent of the error bars in Figs. \ref{fig:groupVel} and~\ref{fig:mixing}.

The measurements of intensity distributions are subjected to inhomogeneities in the detection efficiencies for each of the four internal basis state and inaccuracies in the angles of the statically and dynamically implemented coins. Assuming errors of the detection efficiencies of $\pm 2.5~\%$ and of the coin angles of $\pm 1^{\circ}$, we conduct a Monte Carlo simulation in which we randomly generate 1000 different settings for these quantities within the assumed error range. For each of these settings we calculate the deviation of the resulting numeric intensity distribution from a reference intensity distribution. This reference is obtained when running the numerical simulation with the fit parameters allowing for the closest approximation of the experimental results. The error for the individual positions and polarizations is then calculated as the standard deviation of the randomly generated samples from the reference distribution.

The errors of the similarity to an equidistribution are determined via error propagation from the errors of the intensities, resulting in the error bars visible in Fig. \ref{fig:mixing}.

\end{document}